\documentclass[
    9pt,
    twocolumn,
]{extarticle}

\usepackage[paper=a4paper,top=0.85in,bottom=1.2in,right=0.75in,left=0.75in]{geometry}
\usepackage[utf8]{inputenc}
\usepackage[x11names,dvipsnames]{xcolor}  %
\usepackage[english]{babel}
\usepackage[babel]{csquotes}

\usepackage[
    natbib=true,
    backend=biber,
    style=nature,
    citestyle=numeric-comp,
    sorting=none,
    giveninits=true,
    uniquename=init
]{biblatex}
\usepackage{float}  %
\usepackage{stfloats}  %
\usepackage{graphicx} %
\usepackage[font=small, labelfont=bf]{caption}
\usepackage[]{placeins} %
\usepackage{tikz}

\usepackage{amsfonts}
\usepackage{amssymb}
\usepackage{mathdots}
\usepackage{mathtools}  %
\usepackage{stmaryrd} %

\usepackage{xspace}       %
\usepackage{setspace}     %
\usepackage[normalem]{ulem}         %

\usepackage{footmisc}     %
\usepackage[nobottomtitles]{titlesec}   %
\usepackage[switch]{lineno}             %

\usepackage[noblocks]{authblk}

\usepackage{siunitx}

\usepackage{tabularx}     %
\usepackage{booktabs}     %
\usepackage{diagbox}      %
\usepackage{makecell}     %
\usepackage{multirow}     %
\usepackage{tablefootnote} %
\usepackage{threeparttable}   %

\usepackage{enumitem}

\usepackage[
    colorlinks=true,
    linkcolor=black,
    citecolor=black,
    filecolor=black,
    urlcolor=black,
    plainpages=false,
    pdfpagelabels,
    breaklinks=true,
    ]{hyperref}
\usepackage[capitalize]{cleveref}  %
\crefformat{equation}{Eqn.~#2#1#3}
\Crefformat{equation}{Eqn.~#2#1#3}
\crefrangeformat{equation}{Eqns.~#3#1#4 to~#5#2#6}
\Crefrangeformat{equation}{Eqns.~#3#1#4 to~#5#2#6}
\crefmultiformat{equation}{Eqns.~#2#1#3}{ and~#2#1#3}{, #2#1#3}{ and~#2#1#3}
\Crefmultiformat{equation}{Eqns.~#2#1#3}{ and~#2#1#3}{, #2#1#3}{ and~#2#1#3}

\usepackage[symbols]{glossaries-extra}  %
\glsdisablehyper  %

\setabbreviationstyle[abbreviation]{long-short}
\setabbreviationstyle[acronym]{long-noshort}

\newabbreviation[]{ann}{ANN}{artificial neural network}
    \newcommand{\ann}{\gls{ann}\xspace}
    \newcommand{\anns}{\glspl{ann}\xspace}

\newabbreviation[]{bm}{BM}{Boltzmann machine}

\newabbreviation[]{bp}{BP}{backpropagation}
    \newcommand{\bp}{\gls{bp}\xspace}

\newabbreviation[]{dkl}{\ensuremath{D_\mathrm{KL}}}{Kullback-Leibler divergence}

\newabbreviation[category=acronym]{dtp}{DTP}{(difference) target propagation}

\newabbreviation[]{fa}{FA}{feedback alignment}

\newabbreviation[]{fc}{FC}{fully-connected}

\newabbreviation[]{glm}{GLM}{generalized linear model}

\newabbreviation[]{kp}{KP}{Kolen-Pollack algorithm}

\newabbreviation[]{lif}{LIF}{leaky integrate and fire}

\newabbreviation[]{ng}{NG}{natural gradient}

\newabbreviation[]{mcmc}{MCMC}{Markov chain Monte Carlo}

\newabbreviation[]{pal}{PAL}{phaseless alignment learning}

\newabbreviation[category=acronym]{ppd}{PPD}{phase plane diagram}
    \newcommand{\ppd}{\gls{ppd}\xspace}

\newabbreviation[]{psp}{PSP}{post-synaptic potential}

\newabbreviation[]{rdd}{RDD}{regression discontinuity design}

\newabbreviation[]{sal}{SAL}{spike-based alignment learning}
    \newcommand{\sal}{\gls{sal}\xspace}

\newabbreviation[]{scfa}{SCFA}{sign-concordant feedback alignment}

\newabbreviation[]{ssn}{SSN}{spiking sampling network}
    
    \newcommand{\ssns}{\glspl{ssn}\xspace}
    
\newabbreviation[]{snn}{SNN}{spiking neural network}

\newabbreviation[]{stdd}{STDD}{spike-timing difference distribution}
	\newcommand{\stdd}{\gls{stdd}\xspace}
	\newcommand{\stdds}{\glspl{stdd}\xspace}

\newabbreviation{stdwi}{STDWI}{spike-timing-dependent weight inference}

\newabbreviation[]{wm}{WM}{weight mirror}

\newabbreviation[category=abbrevation]{stdp}{STDP}{spike-timing-dependent plasticity}
    \newcommand{\stdp}{\gls{stdp}\xspace}

\newglossaryentry{convnet}{
    name={Conv\-Net},
    description={ANN part of the SymmNet}
}

\newglossaryentry{rddnet}{
    name={RDD\-Net},
    description={The complementary spiking neural network that implements RDD symmetrization}
}

\newglossaryentry{salnet}{
    name={SAL\-Net},
    description={The complementary spiking network implementing the SAL principle}
}

\newglossaryentry{symmnet}{
    name={Symm\-Net},
    description={Our proposed architecture to test and compare variuous spiking an non-spiking symmetrization algorithms}
}

\glsxtrnewsymbol[description={bla}]{pDeltaT}{\ensuremath{p(\Delta t)}}

\setlist[itemize]{itemsep=0pt, parsep=0pt}

\newcommand{\captionanno}[1]{\textbf{#1)}\nolinebreak\hspace{0.em}\xspace}  %
\newcommand{\captiontitle}[1]{\textbf{#1}}  %
\newcommand{\crefanno}[1]{#1}  %
\newcommand{\figanno}[1]{\textbf{\small#1}}  %
\newcommand{\myemph}[1]{\emph{#1}}  %

\newcommand{\dd}{\ensuremath{\mathrm{d}}}  %
\newcommand{\mat}[1]{\ensuremath{\boldsymbol{#1}}}  %
\newcommand{\var}{\ensuremath{\mathrm{Var}}}  %
\newcommand{\vv}[1]{\ensuremath{\boldsymbol{#1}}}  %
\newcommand{\transp}{\ensuremath{\intercal}}  %

\newcommand{\DeltaT}{\Delta t}  %

\newcommand{\noiseinit}{\ensuremath{\sigma^\mathrm{noise}_\mathrm{syn}}}
\newcommand{\noisestdp}{\ensuremath{\sigma^\mathrm{noise}_\mathrm{plast}}}
\newcommand{\psp}{\kappa}  %
\newcommand{\setReal}{\ensuremath{\mathbb{R}}}  %
\newcommand{\setNat}{\ensuremath{\mathbb{N}}}  %
\newcommand{\spkset}[1]{\left\{ #1 \right\}}  %
\newcommand{\spktrain}{\ensuremath{S}}  %
\newcommand{\stdpAntiCausal}{\alpha_\mathrm{a}}  %
\newcommand{\stdpBoth}{\alpha_\mathrm{a/c}}
\newcommand{\stdpCausal}{\alpha_\mathrm{c}}  %
\newcommand{\stdpwindow}{f}  %
\newcommand{\stdpwindowCausal}{f^+}  %
\newcommand{\stdpwindowAntiCausal}{f^-}  %
\newcommand{\stdpwindowBoth}{f^\pm}  %
\newcommand{\tauref}{\ensuremath{\tau_\mathrm{ref}}}  %
\newcommand{\tausyn}{\ensuremath{\tau_\mathrm{syn}}}  %
\newcommand{\trop}{\ensuremath{T}}  %
\newcommand{\tspk}{\ensuremath{t}}  %

\newcommand{\uP}{\ensuremath{u^\mathrm{P}}}  %
\newcommand{\vuP}{\ensuremath{\vv{u}^\mathrm{P}}}  %
\newcommand{\buP}{\ensuremath{\Bar{u}^\mathrm{P}}}  %
\newcommand{\vuI}{\ensuremath{\vv{u}^\mathrm{I}}}  %
\newcommand{\vuTgt}{\ensuremath{\vv{u}^\mathrm{tgt}}}  %

\newcommand{\vvApi}{\ensuremath{\vv{v}^\mathrm{api}}}  %
\newcommand{\vvBas}{\ensuremath{\vv{v}^\mathrm{bas}}}  %
\newcommand{\bvvBas}{\ensuremath{\Bar{\vv{v}}^\mathrm{bas}}}  %
\newcommand{\bvBas}{\ensuremath{\Bar{v}^\mathrm{bas}}}  %
\newcommand{\vvDen}{\ensuremath{\vv{v}^\mathrm{den}}}  %

\newcommand{\vpsp}{\ensuremath{\hat{\vv{\psp}}}}  %

\newcommand{\lamApi}{\ensuremath{\lambda^\mathrm{api}}}  %
\newcommand{\lamBas}{\ensuremath{\lambda^\mathrm{bas}}}  %
\newcommand{\lamNudge}{\ensuremath{\lambda^\mathrm{nudge}}}  %
\newcommand{\lamDen}{\ensuremath{\lambda^\mathrm{den}}}  %

\newcommand{\WPP}{\ensuremath{W^\mathrm{PP}}}  %
\newcommand{\mWPP}{\ensuremath{\mat{W}^\mathrm{PP}}}  %
\newcommand{\BPP}{\ensuremath{B^\mathrm{PP}}}  %
\newcommand{\mBPP}{\ensuremath{\mat{B}^\mathrm{PP}}}  %
\newcommand{\mWIP}{\ensuremath{\mat{L}^\mathrm{IP}}}  %
\newcommand{\mWPI}{\ensuremath{\mat{L}^\mathrm{PI}}}  %
\newcommand{\Weff}{\ensuremath{W^\mathrm{eff}}}  %
\DeclareMathOperator*{\argmin}{arg\,min}
\DeclareMathOperator{\sgn}{sgn}

\begin{document}

\title{\Huge \bfseries Spike-based alignment learning solves the \\weight transport problem}

\author[1,2]{Timo~Gierlich\thanks{\texttt{timo.gierlich@unibe.ch}}}
\author[2]{Andreas~Baumbach}
\author[2]{Akos~F.~Kungl}
\author[1,3]{Kevin~Max}
\author[1]{Mihai~A.~Petrovici\thanks{\texttt{mihai.petrovici@unibe.ch}}}

\affil[1]{\small Department of Physiology, Bern University, Switzerland}
\affil[2]{\small Kirchhoff-Institut f{\"u}r Physik, Ruprecht-Karls-Universit{\"a}t Heidelberg, Germany}
\affil[3]{\small Neural Computation Unit, Okinawa Institute of Science and Technology, Japan}

\date{}

\maketitle

\begin{refsection}

\begin{abstract}

In both machine learning and in computational neuroscience, plasticity in functional neural networks is frequently expressed as gradient descent on a cost. 
Often, this imposes symmetry constraints that are difficult to reconcile with local computation, as is required for biological networks or neuromorphic hardware. 
For example, wake-sleep learning in networks characterized by Boltzmann distributions assumes symmetric connectivity. 
Similarly, the error backpropagation algorithm is notoriously plagued by the weight transport problem between the representation and the error stream.
Existing solutions such as feedback alignment circumvent the problem by deferring to the robustness of these algorithms to weight asymmetry.
However, they scale poorly with network size and depth.
We introduce spike-based alignment learning (SAL), a complementary learning rule for spiking neural networks, which uses spike timing statistics to extract and correct the asymmetry between effective reciprocal connections.
Apart from being spike-based and fully local, our proposed mechanism takes advantage of noise.
Based on an interplay between Hebbian and anti-Hebbian plasticity, synapses can thereby recover the true local gradient.
This also alleviates discrepancies that arise from neuron and synapse variability -- an omnipresent property of physical neuronal networks.
We demonstrate the efficacy of our mechanism using different spiking network models.
First, SAL can significantly improve convergence to the target distribution in probabilistic spiking networks versus Hebbian plasticity alone.
Second, in neuronal hierarchies based on cortical microcircuits, SAL effectively aligns feedback weights to the forward pathway, thus allowing the backpropagation of correct feedback errors.
Third, our approach enables competitive performance in deep networks using only local plasticity for weight transport. 
\end{abstract}

\section{Introduction}
\label{sec:intro}
\resetlinenumber

Prominent models of neuronal computation rely on core assumptions that inevitably give rise to symmetry constraints on their connectivity. %
For example, prominent recurrent network models such as Hopfield networks \cite{hopfield1982neural} and Boltzmann machines \cite{hinton1983optimal} require a symmetric weight matrix, which also needs to be enforced during wake-sleep learning~\cite{ackley1985learning, hinton2002training}. %
They effectively inherit this property from their spin-glass archetypes in solid-state physics, for which the symmetry of particle interactions follows from fundamental laws of nature.

Perhaps even more prominently, the reverse calculation of gradients in deep neural networks naturally requires knowledge of the forward weights \cite{linnainmaa1976taylor,werbos2005applicationsReprint,rumelhart1986learning}. %
While this weight transport is inconsequential when calculations are simply carried out by an arithmetic logic unit, models of error backpropagation in the brain require the corresponding backward transport circuitry to mirror the forward one \cite{whittington2019theories,richards2019deep,lillicrap2020backpropagation}.

In general, network models for physical neuronal substrates\footnote{
    To make the distinction between \anns as used in deep learning and physical, time-continuous network models for biological or bio-inspired applications, we will refer to the former type as \myemph{neural} networks and to the latter as \myemph{neuronal}.
} 
such as the brain or analog neuromorphic hardware are bound to constraints that are inherent to their physics.
One such restriction is (spatiotemporal) locality, a fundamental property of physical networks which strongly limits the information that may enter synaptic plasticity rules.
Consequently, the physical plausibility of complex, non-local learning algorithms is determined by their amenability to implementation using only local operations.

Another characteristic feature of physical neuronal computing is the inevitable presence of inherent temporal and spatial parameter variations across neurons and synapses.
In the brain, these naturally emerge, for instance, from the morphological variety among different neurons of the same cell type; in analog neuromorphic hardware, component variability inevitably occurs during the manufacturing process.
This expresses the need for homeostatic mechanisms additional to functional learning rules to increase robustness and maintain operational stability~\cite{marder2006variability,watt2010homeostatic,abbott2000synaptic,davis2001maintaining,lee2019mechanisms,yang2023homeostatic}.
A direct and often ignored consequence of substrate heterogeneity is that the effect of the weight on a postsynaptic neuron crucially depends on the neuron parameters (synapse model, conductance values, etc.), meaning that identical weight values elicit different \glspl{psp} in different neurons. 
What counts on the computational level however is the \emph{effective weight}.

On a similar note, the molecular or electronic mechanisms giving rise to synaptic changes are inherently stochastic and variable~\cite{bartol2015nanoconnectomic,noguchi2019bidirectional,durst2022vesicular, Rodrigues2023stochastic,billaudelle2020versatile,atoui2025multitimescale}.
In other words, two different synapses connecting two different neurons will react to the same stimuli by a different amount of long-term potentiation or depression.
This undermines the assumption of perfect weight updates commonly found in models for credit assignment in the brain.

Hence, for promoting functionality and increasing the biological plausibility, learning algorithms for physical neuronal networks should address both the weight transport problem and the resilience to inherent parameter variability.
In the following, we propose a solution to these problems in spiking networks and investigate its effectiveness for two common learning schemes that are notoriously plagued by the weight transport problem: wake-sleep learning~\cite{ackley1985learning, hinton2002training}
and error \bp~\cite{linnainmaa1976taylor,rumelhart1986learning}. 

By describing spiking dynamics as sampling from an underlying Boltzmann distribution, \ssns ~\cite{buesing2011neural,petrovici2016stochastic} create a direct link between spiking neuronal networks and Boltzmann machines.
They are thus able to learn probabilistic internal representations of the world, which offers an algorithmic interpretation of certain activity patterns in the brain~\cite{berkes2011spontaneous,haefner2016perceptual,orban2016neural}, while also enabling the instantiation of Bayesian generative and discriminative models in neuromorphic hardware \cite{kungl2019accelerated,billaudelle2020versatile,czischek2022spiking,klassert2022variational}.
However, they also inherit the weight symmetry constraints from their \ann counterparts.

Additionally, these networks are usually trained with a spiking variant of the contrastive divergence / wake-sleep algorithm \cite{ackley1985learning,hinton2002training,neftci2014event}, which minimizes the difference of correlations between data-constrained and free sampling phases.
Since correlations are symmetric under the exchange of neurons, weight updates of reciprocal synapses must also be identical (as would also follow directly from the general symmetry constraint on the weight matrix).
However, since in physical networks the calculation of these weight updates happens in different synapses that cannot communicate directly, symmetry is not generally guaranteed.
Therefore, the actual weight updates may differ significantly from the ones assumed by the algorithm; apart from deviating from the intended trajectory of learning, this may also lead to a forgetting of previously learned features.
Altogether, this can result in a significant drop in performance, as we also show later.

\begin{figure}[tbp]
    \centering
    \begin{tikzpicture}
        \node[anchor=north west] at (.5, -0.4) {\includegraphics[width=2cm]{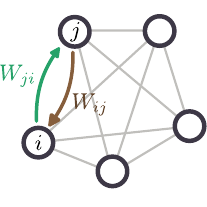}};
        \node[anchor=north west] at (0., 0) {\figanno{a1} \quad\small\textbf{wake-sleep}};
        \node[anchor=north west] at (0.5, -2.5) {\footnotesize$\begin{aligned} \Delta W_{ij} &\propto \langle z_i z_j \rangle_\mathrm{wake} \\  &- \langle z_i z_j \rangle_\mathrm{sleep} \end{aligned}$};
        
        \node[anchor=north west] at (3, -0.5) {\includegraphics[width=2cm]{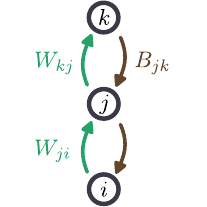}};
        \node[anchor=north west] at (3.3, 0) {\figanno{a2} \quad\small\textbf{backprop}};
        \node[anchor=north west] at (4.7, -1.) {\footnotesize$\begin{aligned} \Delta W_{ji} &\propto e_j r_i \\  e_j &= r_j' B_{jk} e_k \\ B_{jk} &\equiv W_{kj} \end{aligned}$};
        \node[anchor=north west] at (0., -3.7) {\figanno{b}};
        \node[anchor=north west] at (-0.2, -3.7) {\includegraphics[width=7cm]{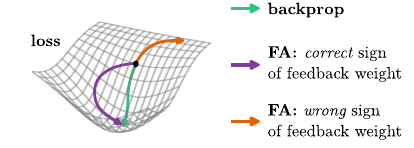}};
        
    \end{tikzpicture}
    \caption{
        \captiontitle{The weight transport problem in physical neuronal networks.}
        Many successful learning rules presuppose certain symmetry properties of the underlying network structure.
        This poses challenges to reconcile them with the locality principle of physical neuronal computation.
        \captionanno{a1} In wake-sleep learning, the network is trained using correlation measurements.
        Since these measurements are carried out by two distinct, individually parameterized synapses (here $W_{ji}$ and $W_{ij}$), weight updates are not symmetric.
        \captionanno{a2} The \gls{bp} algorithm relies on the transportation of the gradient across layers, which requires a copy of the forward weights (green) to the backward path (blue).
        \captionanno{b} Depending on the misalignment between the true \gls{bp} gradient (green arrow) and the trajectory followed by \gls{fa}, learning can be slowed down significantly (violet) or fail completely (orange).
    }
    \label{fig:intro}
\end{figure}

Similar issues have initially led to a strong pushback against \bp-like learning in the brain \cite{crick1989recent,lillicrap2020backpropagation}.
With the emergence of biologically plausible adaptations of \gls{bp} 
\cite{lansdell2019learning,ernoult2022scaling,xie2003equivalence,scellier2017equilibrium,sacramento2018dendritic,pozzi2020attention,kolen1994backpropagation,payeur2021burstdependent,roelfsema2005attention,pozzi2018biologically,haider2021latent,ellenberger2024backpropagation,senn2024neuronal,lee2015difference}, several issues of standard \bp have been mitigated;
however, many of these algorithms still (at least implicitly) rely on copying the weights from the bottom-up pathways to the top-down pathways for correct transportation of errors; for example, approaches such as the Kolen-Pollack algorithm and variants~\cite{kolen1994backpropagation,akrout2019deep,roelfsema2005attention,pozzi2018biologically} defer the weight transport problem to a \textit{weight update} transport problem by assuming identical weight updates in the forward and backward path -- an unrealistic premise in the presence of noise.
Another common way to circumvent the weight transport problem is to ignore it altogether.
\Gls{fa} \cite{lillicrap2016random} builds on the observation that during training, forward weights tend to align to random, but fixed backward connections, and can therefore transport meaningful errors across layers.
However, \gls{fa} is known to scale poorly in deeper networks~\cite{bartunov2018assessing,moskovitz2018feedback,max2024learning} (see also \cref{fig:intro}\crefanno{b}).

Furthermore, only few bio-plausible adaptations of \gls{bp} consider networks of spiking neurons, e.g.~\cite{burbank2015mirrored,guerguiev2017deep,payeur2021burstdependent}.
They all have in common that they require additional mechanisms for weight transport. 
Recently developed models
such as \gls{rdd}~\cite{guerguiev2019spikebased} and \gls{stdwi}~\cite{ahmad2020overcoming} propose to infer the strength of the forward weights through a specific stimulation of the corresponding pre- and postsynaptic neurons, which introduces distinct symmetrization phases during learning.
This work fills fills the gap by providing a fully spike-based solution to the weight transport problem, that can, in principle, be applied concurrently during functional learning, similar to its rate-based sibling \gls{pal} \cite{max2024learning}.
In addition, we also introduce a bio-plausible, spiking implementation of error~\gls{bp} based on dendritic cortical microcircuits~\cite{sacramento2018dendritic,haider2021latent}.

In summary, physically plausible realizations of functionally powerful learning rules require additional homeostatic mechanisms to establish robustness against inevitable parameter variability across neurons and synapses and to maintain operational stability.%
In this work, we address these problems and propose that nature has found a solution to the challenges of weight transport and robustness in noisy inhomogeneous substrates.
We demonstrate that a particular form of \stdp is capable of aligning the synaptic weights of reciprocally connected neurons in a recurrently connected network.
We therefore call this framework \gls{sal}.
Similarly to \gls{pal}, \gls{sal} is designed to augment existing spike-based learning rules that are subject to the weight transport problem.

To this end, we combine three essential, but otherwise basic elements of cortical dynamics.
We leverage the \emph{spike-based communication} between neurons in order to directly access temporal discrepancies arising from asymmetries in the effective connectivity.
However, these only become tractable in the presence of \emph{noise}, here an explicit feature rather than a bug.
Ultimately, the resulting information can enter a modified \emph{\stdp rule} which thereby becomes capable to correct undesirable deviations.

\section{Results}
\label{sec:results}

\resetlinenumber

\subsection{Neuron and synapse model}
\label{sec:results:model}

\begin{figure*}[tb]
    \centering
    \begin{tikzpicture}
   
        \node[anchor=north west] at (0.25, 0) {\includegraphics[width=4.5cm]{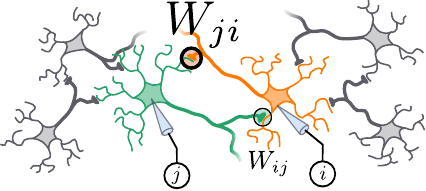}};
        \node[anchor=north west] at (0.25, -2.1) {\includegraphics[width=4.6cm]{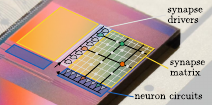}};
        \node[anchor=north west] at (0, 0) {\figanno{a}};
    
        \node[anchor=north west] at (0.75, -4.7) {\includegraphics[width=3.9cm]{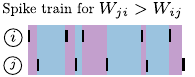}};
        \node[anchor=north west] at (0, -4.7) {\figanno{b}};
    
        \node[anchor=north west] at (5.7, 0) {\includegraphics[width=4.4cm]{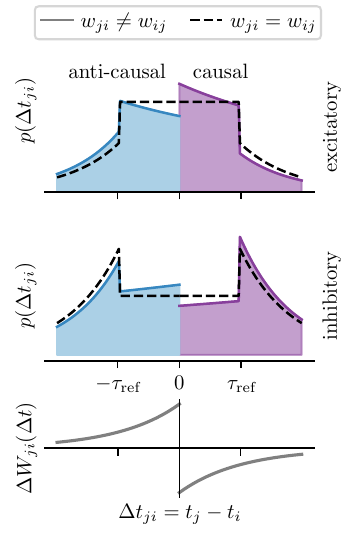}};
        \node[anchor=north west] at (5.2, 0) {\figanno{c1}};
    
        \node[anchor=north west] at (5.2, -4.7) {\figanno{c2}};
        \node[anchor=north west] at (10.4, -3.3) {\includegraphics[width=5.4cm]{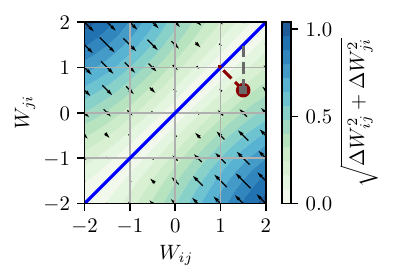}};
        \node[anchor=north west] at (10.3, -3.5) {\figanno{e}};
    
        \node[anchor=north west] at (10.6, 0.1) {\includegraphics[width=4.9cm]{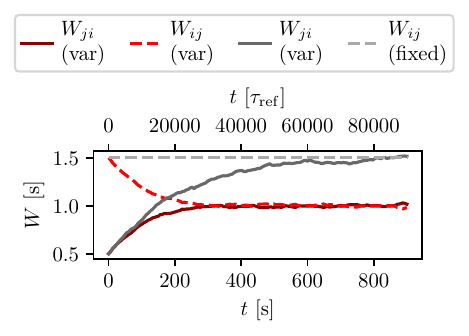}};
        \node[anchor=north west] at (10.3, -0.0) {\figanno{d}};
    \end{tikzpicture}
    \caption{\captiontitle{Principles of spike-based alignment learning.}
    \captionanno{a} Two reciprocally connected neurons $i$ and $j$ embedded in a recurrent neural network fire stochastically due to external spike noise.
    Here, weight $W_{ji}$ (orange synapse) is stronger than $W_{ij}$ (green).
    Because \gls{sal} symmetrizes \emph{effective} weights rather than pure synaptic strengths, effects of morphological differences between the two neurons or the synaptic location at the dendritic tree are implicitly balanced out.
    This underpins the feasibility of \sal in both biological and neuromorphic substrates (illustration adapted from \cite{mueller2020bss2}).
    \captionanno{b} Example spike trains with colored spike-timing differences $\DeltaT$ (causal in purple and anti-causal in blue), as seen from the perspective of $W_{ji}$.
    \captionanno{c1} \Gls{stdd} distribution for two reciprocally connected neurons as described in \cref{sec:results:sal_spiketiming}.
    The upper panel shows two excitatory weights, the lower two inhibitory ones.
    \captionanno{c2} \Gls{stdp} with an anti-Hebbian window used by \gls{sal} for weight alignment.
    \captionanno{d} Time course of symmetrization with \gls{sal}.
    If both weights are plastic, they converge to their common mean (red); if only one weight uses \sal, it converges to the other one (gray).
    \captionanno{e} Phase diagram showing the evolution of the two weights under \gls{sal}.
    The arrows indicate the direction of the weight update through \gls{sal}, the color map in the background the magnitude of the update.
    The blue line indicates the attractor of \gls{sal} which lies on the diagonal $W_{ij} = W_{ji}$.
    \Gls{sal} always converges to the desired solution $W_{ij} = W_{ji}$ from any starting point in the $W_{ij}$-$W_{ji}$-plane.
    The example trajectories from d) are depicted in red and gray.
    }
    \label{fig:results:sal}
\end{figure*}

Consider two neurons $i$ and $j$  that are mutually connected through the weights $W_{ij}$ and $W_{ji}$ and are affected by different types of spatial and temporal noise (see \cref{fig:results:sal}\crefanno{a}).
On analog neuromorphic chips, hardware components are subject to inevitable variability in the manufacturing process, the so-called fixed pattern noise, that cannot be fully compensated by calibration~\cite{petrovici2014characterization,pehle2022brainscales,grubl2020verification}.
Likewise, neurons of the same type vary in their morphology such as cell body size and dendritic tree structure, as well as physiological properties such as ion channel density~\cite{kawaguchi1993groupings,liu2024neuronal}.
These factors critically influence the input-output relationship of different neurons~\cite{mednikova2020heterogeneity,rathour2022voltage}. 
Hence, what counts from a computational point of view is not so much a parameter of the synapse alone, but rather the \emph{effective weight}, i.e., the effect of a synaptic event on the postsynaptic firing probability, which is what we describe as $W$.

Temporal noise is also omnipresent in any physical substrate.
The list of sources include thermal noise, sensory noise or the stochastic nature of ion channels and receptors~\cite{faisal2008noise}.
Furthermore, cortical neurons are known to undergo constant bombardment with irregular spike trains, some of which may be considered background noise~\cite{brunel2000dynamics,fourcaud2002dynamics,jordan2019deterministic,dold2019stochasticity}.
The situation in neuromorphic hardware is similar, with noise on neuronal membranes being either intrinsic (for analog neurons) or extrinsic, through background synaptic bombardment.
We thus consider neurons to operate in a regime of stochastic spiking.

For mathematical tractability, we choose a \gls{glm} as our neuron model, although the working principle of \gls{sal} is of a more general nature and  thus applicable to other models as well.
Each neuron is described by its membrane potential $u_i$, 
\begin{equation}\label{eq:results:mempot}
    u_i(t) = b_i + \sum_{k} W_{ik}(t) \int \limits_{0}^{\infty} \psp(s) S_k(t - s) \dd s,
\end{equation}
where $b_i$ is a constant bias, $W_{ik}$ are the input weights, $\psp(t)$ the kernel of the \gls{psp}
and $S_k(t)$ the input spike train from neuron $k$ (for more details see \cref{sec:methods:neuronmodel}).

The noise sources and resulting voltage fluctuations are not modeled explicitly, instead we capture the emerging randomness in the inherent probabilistic spiking mechanism of the \gls{glm}.
For simplicity, we assume that the noise process has a constant mean  over time, which simply enters the neuronal bias $b_i$, along with other constant biases such as the leak potential.
The output spikes are produced by an inhomogeneous Poisson process with an absolute refractory period of length $\tauref$, where the instantaneous firing probability of a non-refractory neuron is given by $r_i(t) = \tauref^{-1} \exp \left( u_i(t) \right)$.

\subsection{Spike timing correlations reflect weight asymmetry}
\label{sec:results:sal_spiketiming}

The working principle of \gls{sal} resides on the observation that weights leave a characteristic imprint on the cross-correlation between pre- and postsynaptic spike trains.
Each spike generated by one of the neurons elicits a \gls{psp} in the respective postsynaptic partner. For an excitatory synapse, this transiently raises the postsynaptic neuron's firing probability.
The magnitude of the change is dependent on the synaptic weight, meaning that the postsynaptic neuron reacts on average more often, and especially earlier, to an incoming spike for a larger weight.
Correspondingly, a spike transmitted through an inhibitory synapse would create a negative \gls{psp}, causing the postsynaptic neuron to fire less often and later.
Hence, the distribution of the spike timing difference $\DeltaT_{ij} = t_j - t_i$ carries information about the synaptic weights.
In the following, we will call this the \glsxtrfull{stdd} $p_{ij}(\DeltaT)$.
Only nearest-neighbor pre-post spike pairs are taken into account, since the timing of the first postsynaptic spike carries all relevant information.

To illustrate the effects of weight asymmetry on the \stdd, consider the case of two neurons reciprocally connected by excitatory weights with $W_{ji} > W_{ij} > 0$.
	For clarity, the curves in \cref{fig:results:sal} are calculated for rectangular \glspl{psp}, and a detailed discussion of other kernel shapes is provided in \cref{sec:results:other_psps}.

Every time neuron $i$ spikes, neuron $j$ will \enquote{answer} with a certain probability with a postsynaptic spike after some $\DeltaT$, just as neuron $j$ will respond to a spike from $i$ \cref{fig:results:sal}\crefanno{b}.
However, since $W_{ji}$ is greater than $W_{ij}$, neuron $j$ will respond to neuron $i$ on average faster than vice versa.
Therefore, the probability density is shifted on the right-hand side of the \gls{stdd} (causal from the point of view of $W_{ji}$, $\DeltaT_{ji} > 0)$ towards the center, and on the left-hand side (anti-causal ones, $\DeltaT_{ji} < 0$) away from the center to the left (\cref{fig:results:sal}\crefanno{c1}, top panel).
This leads to an asymmetry between the left and right side of the \gls{stdd}, which is informative of the weight difference.
Importantly, both synapses observe the same spikes and therefore the \gls{stdd} of $W_{ij}$ is the same as that of $W_{ji}$ but mirrored at $\DeltaT=0$.
Thereby, weight information of both synapses is directly available at each synapse and can be used for a local \gls{stdp}-based plasticity rule.

The effect of excitatory reciprocal PSPs on the \gls{stdd} can thus be summarized as follows.
First, the higher the average value of the weights is, the more probability mass is concentrated in the interval between $-\tauref$ and $\tauref$, and correspondingly less mass is contained in the tails of the distribution.
Second, the greater the difference between the two weights, the more mass is transported towards $\DeltaT=0$ on the causal half of the distribution and away from it on the anti-causal half (from the perpective of the weight that is too strong), which increases the asymmetry around $\DeltaT=0$.

The same principles hold if one or both synapses are inhibitory, with the only difference being that a strong inhibitory synapse pushes the probability of a post-synaptic spike to larger $\DeltaT$.
In general, we can thus conclude the following.
First, the smaller the average of the two weights, the more mass is located in the tails of the distribution.
Second, the greater the difference between two weights, the more mass is transported away the y-axis on the causal half of the distribution and towards it on the anti-causal half.

The key benefit of having noise in the system becomes apparent here.
Without noise, single synapses might be too weak to contribute significantly to the \gls{stdd}, while strong synapses would pull its entire mass towards zero.
Noise allows a smooth sampling of all possible $\DeltaT$, and allows all synapses to contribute to it.

\subsection{Spike-based alignment learning}
\label{sec:results:sal_definition}

As synapses have direct, local access to the \gls{stdd}, they can use it to correct asymmetries towards their otherwise inaccessible reciprocal counterparts.
Just like classical \stdp harnesses the \stdd for Hebbian learning, \sal uses it for symmetrization.
	
Instead of a classical Hebbian \gls{stdp} window with a positive causal and a negative anti-causal branch, \sal uses an anti-Hebbian window (\cref{fig:results:sal}\crefanno{c2}):
\begin{equation}\label{eq:results:stdp}
    \Delta W_{ij} = \begin{cases}
        \eta\, \stdpCausal\, \stdpwindowCausal(\DeltaT_{ij}) \quad \text{if } \DeltaT_{ij} \ge 0 \ \text{(causal),}\\
        \eta\, \stdpAntiCausal\, \stdpwindowAntiCausal(\DeltaT_{ij}) \quad \text{if } \DeltaT_{ij} < 0 \ \text{(anti-causal),}\\
    \end{cases}
\end{equation}
Here, $\eta$ is the learning rate and $f(\DeltaT)$ describes the shape of the \stdp kernel, which we assume to be $\stdpwindowBoth(\DeltaT ) = \exp \left( \mp\DeltaT / \tauref \right)$ unless stated otherwise.
The prefactors $\stdpBoth$ determine the type of plasticity, such as Hebbian, anti-Hebbian or \sal.
In \sal, they are chosen such that causal spike pairs weaken the synapse ($\stdpCausal = -1$) and anti-causal ones strengthen it ($\stdpAntiCausal = 1$).
This simple plasticity rule is capable of extracting the asymmetry from the \gls{stdd} to produce average weight updates that align the two weights such that they converge to their mean.
\Cref{fig:results:sal}\crefanno{d} shows the result of a numerical simulation for a pair of reciprocally connected neurons; a detailed analytical proof of the stability of fixed points under \sal is given in \cref{sec:methods:proof}.

The shape of the \gls{stdp} window $f$ plays a key role in extracting the information relevant for symmetrization from the \gls{stdd}:
For instance, an exponential window $f(\DeltaT) = \exp(\DeltaT / \tau)$ \enquote{looks} primarily at $\DeltaT$-values close to zero, where the weight differences manifest themselves as asymmetry around $\DeltaT=0$.
The \gls{stdp} time constant should roughly match the width of the \gls{psp}-kernel, because it determines the typical size of features in the \gls{stdd}.

Importantly, \sal can be applied asymmetrically to reciprocal synapses.
For instance, it is sufficient for only one synapse to be equipped with a \sal rule in order for it to follow its reciprocal weight (\cref{fig:results:sal}\crefanno{d,e}, gray).
This is particularly useful in setups with distinct forward and backward streams, such as for \gls{bp}, where the forward weights are learned with an error-correcting learning rule and the backward weights use \sal for alignment (see \cref{sec:results:mc}).
As shown in the \gls{ppd} \cref{fig:results:sal}\crefanno{e}, \gls{sal} is able to symmetrize weights over a large parameter range.

Conceptually, %
the interplay between functional plasticity and \sal can be understood as follows.
Under theoretically ideal circumstances (no noise, perfect initialization), networks that require weight symmetry would start out with reciprocal synaptic weights already lying on the diagonal of the \ppd, and functional plasticity would move these weights along the diagonal.
In realistic scenarios, reciprocal weights would be more randomly distributed, and functional plasticity would, in general, not drive them towards symmetry, and maybe even away from it.
With \sal, plasticity receives a persistent, orthogonal drive towards its stable manifold on the diagonal, enabling the functional component to operate correctly along its orientation.

Importantly, just like classical Hebbian \stdp, \gls{sal} only uses information available at the locus of the synapse, and a plasticity kernel compatible with observations from human cortex \cite{koch2013hebbian}.
This makes it a suitable candidate for how nature might have addressed the weight transport problem.
The learning rule is also compatible with implementations of \gls{stdp} on neuromorphic platforms \cite{pfeil2013six,serrano2013stdp,qiao2015reconfigurable,davies2018loihi,pehle2022brainscales}, which alleviates the problem of fixed-pattern noise in analog systems and avoids expensive copy operations in digital ones.

\subsection{SAL in functional spiking networks}
\label{sec:results:experiments}
\resetlinenumber

\gls{sal} is designed to enhance functional learning rules that by definition require weight symmetry and are therefore difficult to reconcile with the noisy reality of physical neuronal substrates, whether biological or artificial.
\gls{sal} can be implemented either in a phaseless manner or by adding reoccurring symmetrization phases during learning; in programmable hardware, such phases are easily introduced, while in biology, they can occur while the brain is not attending to external sensory stimuli, most prominently during sleep.

In the following, we present three scenarios that demonstrate the effectiveness of \gls{sal}:
First, we turn to \glspl{ssn} \cite{buesing2011neural, petrovici2016stochastic}, which require both initial weight symmetry and symmetric weight updates for exact gradient descent.
We show that \glspl{ssn} are sensitive to both noise on the initial weight matrix and on the \gls{stdp} mechanism and how \gls{sal} improves learning under such inhomogeneous conditions.

Second, we demonstrate the efficacy of \gls{sal} in a microcircuit model of hierarchical cortical computation \cite{sacramento2018dendritic, haider2021latent}. 
There, the local error correction relies on the backward transportation of gradients through cortical microcircuits, which inherits the weight transport problem from classical error backpropagation.
We show that the microcircuits equipped with \gls{sal} evolve much more similarly to \gls{bp} compared to \gls{fa}, which is prone to misalignment.

Third, we apply \gls{sal} in a deep convolutional network trained on standard image classification tasks to provide evidence that our approach scales to large and deep architectures, while also benchmarking it against other spiking and non-spiking symmetrization algorithms.

\subsubsection{SAL in spiking sampling networks}
\label{sec:results:ssn}

\begin{figure}[t!]
    \centering
    \includegraphics[width=0.85\linewidth]{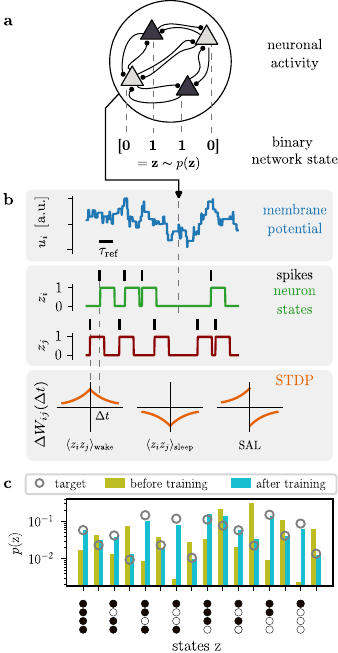} 

    \caption{
        \captiontitle{Working principle of \glspl{ssn}.}
        \captionanno{a} An \gls{ssn} consists of a recurrent spiking network with symmetric reciprocal connections in theory, but asymmetric ones in practice. 
        \captionanno{b} Each neuron fires stochastically as a function of the membrane potential (blue), which gives rise to a sampling process from an underlying distribution $p(\vv z)$.
        The refractory state of each neuron is mapped to a binary variable $z \in [0, 1]$ (green and red).
        \Gls{stdp} with a left-right symmetric window (orange) is used to implement spike-based wake-sleep.
        Because the synaptic update is local to each synapse, reciprocal weight updates are also asymmetric.
        \captionanno{c} Example sampled distribution for $N=4$ neurons.
    }
    \label{fig:results:ssn_principle}
\end{figure}

\begin{figure*}[htbp]
    \centering
    \begin{tikzpicture}
        \node[anchor=north west] at (0.0, -2.0) {\includegraphics[width=1.5cm]{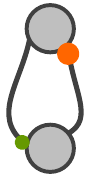}};
        \node[anchor=north west] at (2.0, 0.0) {\includegraphics[width=14.0cm]{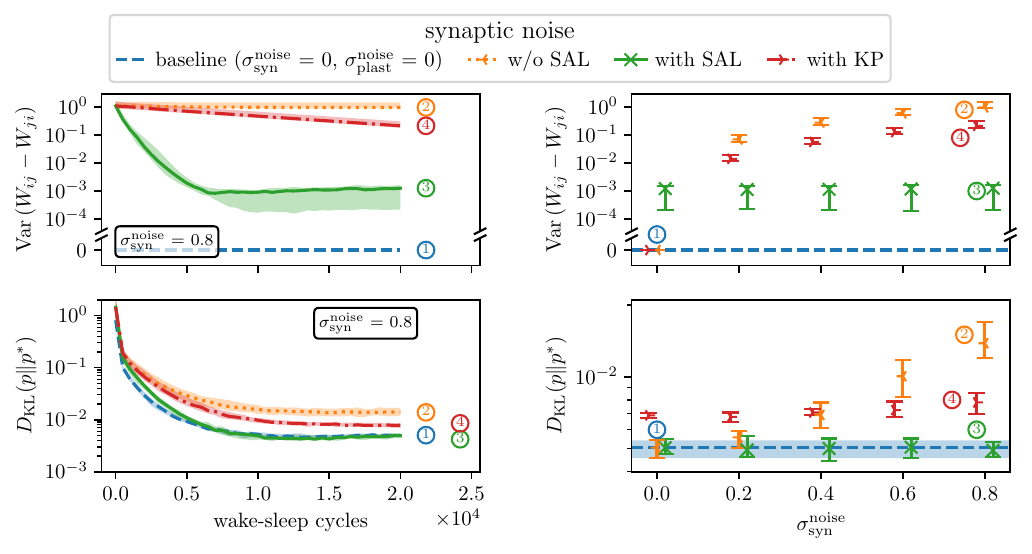}};
        \node[anchor=north west] at (2.0, -1.2) {\figanno{a1}};
        \node[anchor=north west] at (2.0, -4.4) {\figanno{b1}};
        \node[anchor=north west] at (9.0, -1.2) {\figanno{a2}};
        \node[anchor=north west] at (9.0, -4.4) {\figanno{b2}};
    \end{tikzpicture}
    \vspace{-0.25cm}
    \caption{\captiontitle{Synaptic noise in \ssns.}
    An \gls{ssn} is initialized with a symmetric weight matrix and trained with pure \gls{stdp}-based wake-sleep (blue curve), which serves as a baseline (no noise).
    We also train a noised weight matrix (additive Gaussian noise with standard deviation \noiseinit) with (green) and without (orange) a \gls{sal} phase in addition to \gls{stdp}-based wake-sleep learning and the \glsxtrlong{kp}~(\glsxtrshort{kp}, red).
    \captionanno{a1} Evolution of the variance of weight differences $W_{ij} - W_{ji}$ as a measure for the weight asymmetry for the three cases after each epoch.
    \captionanno{a2} Variance of the weight differences after training.
    \captionanno{b1} Evolution of the \gls{dkl} during training between the freely sampled model distribution $p$ and the target distribution $p^*$. 
    \captionanno{b2} Final \gls{dkl} after training as a function of the initial noise variance.
    Note that we plotted the otherwise potentially overlapping values in a2 and b2 slighty staggered in x-direction to increase readability; the orange markers represent the correct x-value.
}
    \label{fig:results:snn_init_noise}
\end{figure*}

\begin{figure*}[htbp]
    \centering
    \begin{tikzpicture}
        \node[anchor=north west] at (0.0, -1.) {\includegraphics[width=2.0cm]{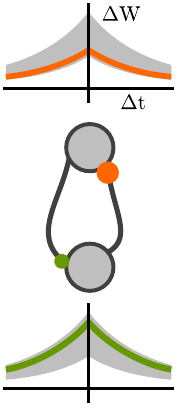}};
        \node[anchor=north west] at (2.5, 0.0) {\includegraphics[width=14.0cm]{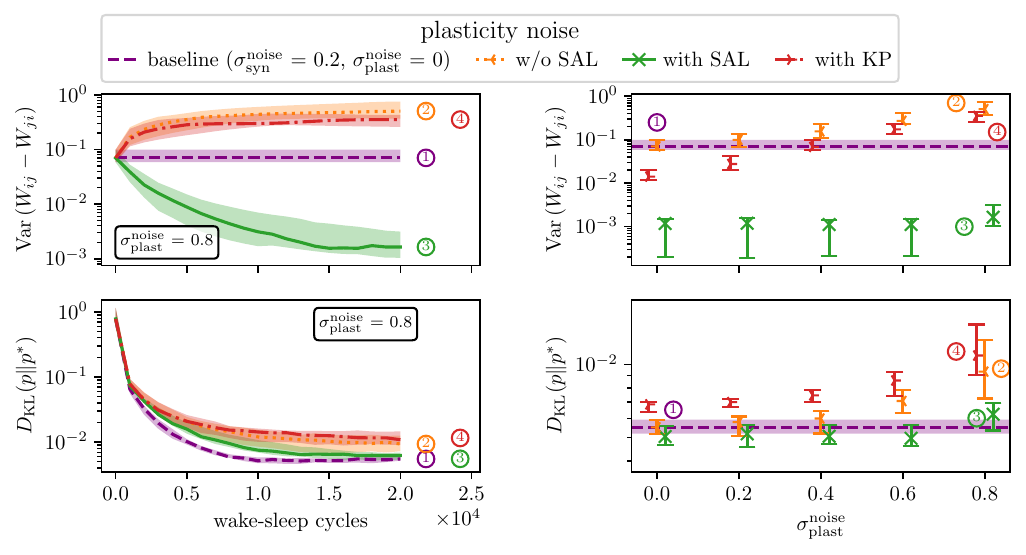}};
        \node[anchor=north west] at (2.5, -1.2) {\figanno{a1}};
        \node[anchor=north west] at (2.5, -4.4) {\figanno{b1}};
        \node[anchor=north west] at (9.5, -1.2) {\figanno{a2}};
        \node[anchor=north west] at (9.5, -4.4) {\figanno{b2}};
    \end{tikzpicture}
    \vspace{-0.25cm}
    \caption{\captiontitle{Plasticity noise in \ssns.}
    The basic configuration of \cref{fig:results:snn_init_noise} is repeated.
    In addition to an initial weight noise of $\noiseinit=0.2$, we now also add noise $\noisestdp$ to the \gls{stdp} kernel.
    For comparison, we show a baseline with $\noiseinit = 0.2$, but without \gls{stdp} kernel noise.
    \captionanno{a1} Evolution of the weight asymmetry $\var(W_{ij} - W_{ji})$ for the three cases (noise without symmetrization, orange; noise with \gls{sal}, green and noise with \glsxtrlong{kp}~(\glsxtrshort{kp}, red).
    \captionanno{a2} Weight asymmetry after training as a function of the~\gls{stdp} noise amplitude.
    \captionanno{b1} Evolution of the \gls{dkl} during training.
    \captionanno{b2} \gls{dkl} after training as a function of~$\noisestdp$.
    In all panels, we report the median and interquartile range over 20 different seeds.
    As in \cref{fig:results:snn_init_noise}, we plotted the values in a2 and b2 slighty staggered in x-direction to increase readability; the orange markers represent the correct x-value.
    }
    \label{fig:results:ssn_stdp_noise}
\end{figure*}

In this section, we demonstrate the effectiveness of \gls{sal} in the framework of~\glspl{ssn}~\cite{buesing2011neural,petrovici2016stochastic}.
As a model for the Bayesian brain, \glspl{ssn} can explain characteristics of perception in noisy, ambiguous environments~\cite{pouget2013probabilistic,fiser2010statistically}.
These networks form probabilistic latent representations of the world and explore these learned internal states through spontaneous activity (sampling).
Interestingly, the sampling process is driven by noise, the same resource that \gls{sal} uses for synaptic symmetrization.
Due to the complex correlations that they need to learn between their constituent neurons, \glspl{ssn} provide a powerful test case for \gls{sal}.

\Glspl{ssn} represent a natural way to mathematically formalize the dynamics of sampling in a biologically plausible manner by extending the theory of Boltzmann machines to spiking networks.
However, they also inherit the need for a symmetric weight matrix from their machine learning counterparts.

An \gls{ssn} consists of $N$ bidirectionally connected neurons, each modeled as a \gls{glm} following \cref{eq:results:mempot}, with absolute refractory time $\tauref$, rectangular \glspl{psp}, and a logistic activation function.
Each neuron $k$ is assigned a binary state: $z_k = 1$ if the neuron is refractory and $z_k = 0$ otherwise, resulting in a network state vector $\vv{z} \in [0, 1]^N$ (see \cref{fig:results:ssn_principle}).
The probability for the network to be in a state $\vv{z}$ is given by
\begin{equation}\label{eq:results:bm_distr}
    p(\vv{z}) = \frac{1}{Z} \exp \left[ \frac{1}{2} \vv{z}^{\transp} \vv{W} \vv{z} + \vv{z}^{\transp}\vv{b}\right],
\end{equation}
where $Z = \sum_{\vv{z}} \exp \left[ \frac{1}{2} \vv{z}^\transp \vv{W} \vv{z} + \vv{z}^\transp\vv{b}\right]$ represents the normalization factor, $\vv{W} \in \setReal^{N \times N}$ is a symmetric weight matrix (i.e., $\vv{W} = \vv{W}^\transp$) and $\vv{b} \in \setReal^{N}$ a vector containing all neuronal biases. 

The network can be trained to a suitable target distribution $p^*(\vv{z})$ by gradient descent on the Kullback-Leibler divergence between $p$ and $p^*$.
This ultimately yields the wake-sleep algorithm with a local Hebbian learning rule~\cite{ackley1985learning}.
Training consists of two alternating phases: a wake phase, which is constrained by the target distribution, and a sleep phase, in which the network samples freely from its current internal distribution.
A weight update is then given by
\begin{equation}\label{eq:results:ws_weights}
    \Delta W_{ij} = \eta \left[ \langle z_i z_j \rangle_\mathrm{wake} - \langle z_i z_j \rangle_\mathrm{sleep}\right],
\end{equation}
where $\eta$ denotes the learning rate and $\langle \cdot \rangle_x$ the expectation value in the respective phase.
Note that because of $\langle z_i z_j \rangle = \langle z_j z_i \rangle$, wake-sleep will always produce symmetric weight updates.

In the framework of \glspl{ssn}, the estimation of $\langle z_i z_j \rangle_x$ can be carried out by \gls{stdp} as introduced in \cref{eq:results:stdp}:
We use left-right-symmetric, triangular \gls{stdp} windows with $f(\DeltaT) = \max\left(-\frac{\DeltaT}{\tauref} - 1, 0\right)$ and $\stdpBoth =  1$ for the wake and $\stdpBoth =-1$ for the sleep phase to get an approximation of $\langle z_i z_j \rangle_x$ purely by observing the spikes.
A global gating signal can induce the phase switching~\cite{neftci2014event}.
To distinguish between \cref{eq:results:ws_weights} and its bio-plausible
version relying on \gls{stdp}, we will refer to the former as \myemph{state-based} wake-sleep and to the latter as \myemph{spike-based} wake-sleep;
throughout this work, we use spike-based wake-sleep learning.
Because of synapse-specific parameter variations in physical substrates, spike-based wake-sleep can be subject to noise, which we model as synapse-specific variations of $\stdpBoth$.
The biases are trained with
$\Delta b_i = \eta \left[ \langle z_i \rangle_\mathrm{wake} - \langle  z_i \rangle_\mathrm{sleep} \right]$, 
which boils down to a measurement of the firing rate.

When integrating the \gls{ssn} model with the \gls{stdp} version of wake-sleep on noisy physical substrates, we encounter the challenges discussed in \cref{sec:intro}:
First, the network must be initialized with symmetric weights, which requires the transport of effective weights between synapses.
Any initial weight asymmetry will persist if symmetric weight updates are applied.
Second, the \gls{stdp} mechanism for the estimation of $\langle z_i z_j \rangle$ is located at a specific synapse $W_{ij}$ and is subject to parameter variations that are, in general, not identical to those of $W_{ji}$.
Therefore, the \gls{stdp} measurements for $\langle z_i z_j \rangle$ and $\langle z_j z_i \rangle$ will differ and thereby violate the requirement for symmetric weight updates during wake-sleep. 
Asymmetric weight updates will lead to diverging weight pairs, which distort the learned distribution and can even unlearn previously learned representations.

We address these issues by introducing an additional \sal update that co-occurs during the sleep phase. %
\Gls{sal} quickly levels initial weight asymmetry and facilitates learning in noisy \gls{stdp} scenarios by counterbalancing weight divergence. 
This stabilizes learning, particularly in lifelong online learning settings.
It removes any assumption of implicit information exchange between synapses, ensuring that wake-sleep is truly local.

To illustrate the effect of \gls{sal}, we conduct two types of experiments, in which we consecutively introduce different types of noise: the \emph{synaptic noise scenario} and the \emph{plasticity noise scenario}.
In both scenarios, we train a network of $N=7$ neurons to approximate a target Boltzmann distribution~$p^*$.

For comparison, we also simulate the training with the \gls{kp} \cite{kolen1994backpropagation} based on weight decay, 
\begin{equation}\label{eq:results:kp}
    \Delta W_{ij} = \eta \left[ \langle z_i z_j \rangle_\mathrm{wake} - \langle z_i z_j \rangle_\mathrm{sleep}\right] - \lambda W_{ij},
\end{equation}
where $\lambda > 0$ represents the weight decay rate.
We dedicate \cref{sec:sm:kp_lr} to an in-depth examination of the effect of $\lambda$ in \glspl{ssn} and how to choose them.

\paragraph{Synaptic noise scenario:}

The first experiment (\cref{fig:results:snn_init_noise}) simulates fixed pattern noise on the synaptic weights by randomizing the initial weight matrix.
To do so, we add Gaussian noise with variance $\noiseinit$ to the weight matrix.
All synapses share identical \gls{stdp} parameters ($\stdpBoth =\pm 1$), making the learning process equivalent to state-based wake-sleep.

\Gls{sal} demonstrates a rapid capability to symmetrize weights, and is able to recover optimal training performance~(\cref{fig:results:snn_init_noise}\crefanno{a1}).
This is consistent across various noise levels and weight differences (\cref{fig:results:snn_init_noise}\crefanno{a2}).
The ability of an \gls{ssn} to learn target distributions is found to be critically dependent on the symmetry of reciprocal weights.
Even minor deviations from symmetry hinder learning (\cref{fig:results:snn_init_noise}\crefanno{b1,b2}).
By applying \gls{sal} simultaneously to the sleep pahse, we demonstrate that it does not impede learning speed (\cref{fig:results:snn_init_noise}\crefanno{b1}).
Consequently, \gls{sal} enables recovery of training performance comparable to the noise-free baseline (\cref{fig:results:snn_init_noise}\crefanno{b2}, blue). 

For \gls{kp}, we choose the weight decay rate that yields the best \gls{dkl} for a noise level of 0.8 (see \cref{fig:sm:kp_lr}).
Still, \gls{sal} yields better results both in terms symmetrization and \gls{dkl} (see \cref{fig:results:snn_init_noise}\crefanno{a2} and \crefanno{b2}).
For small noise levels $\sigma_\mathrm{syn}^\mathrm{noise} < 0.4$, \gls{kp} actually worsens the \gls{dkl} compared to wake-sleep without any symmetrization at all.
This is effectively due to the type of unlearning that \gls{kp} invariably introduces through the weight decay.
Since wake-sleep weight updates are gradually reduced during training as the sampled and target distributions become more similar, the weight decay plays an increasingly dominant role, effectively preventing fine-tuning.
This does not happen with \gls{sal}, as it always conserves an average\footnote{
    More generally in heterogenous systems, where the \gls{sal} learning rates of the two weights are not identical, \gls{sal} updates point towards the \emph{weighted} average $\bar{w}_{ij} = \bar{w}_{ji} = \frac{\eta_{ij}w_{ij} + \eta_{ji}w_{ji}}{\eta_{ij} + \eta_{ji}}$, where $\eta_{ij}$ denotes the individual learning rate of synapse $ij$.
} of the reciprocal synaptic weights.
In our simulations, the size of functional weight updates decreases in the course of the training, such that it equilibrates with the weight decay term at some point. %
From that point onward, training stagnates, leading to poorer performance compared to \gls{sal}, in which wake-sleep and symmetrization is completely decoupled.

\paragraph{Plasticity noise scenario:}
The second experiment (\cref{fig:results:ssn_stdp_noise}) introduces noise to both the initial weight matrix and the wake-sleep \stdp kernels.
I.e., an initial noise value of $\noiseinit =0.2$ is used;
additionally, a random variable~$\xi$ is drawn from a truncated Gaussian distribution for each synapse to model \gls{stdp} inhomogeneity ($\stdpBoth =\pm 1\pm \xi$).

Even small discrepancies in \gls{stdp} parameters for wake-sleep lead to weight divergence in reciprocal synapses (\cref{fig:results:ssn_stdp_noise}\crefanno{a1}).
\Gls{sal} effectively counteracts the tendency of reciprocal weights to diverge, maintaining alignment throughout the training process (\cref{fig:results:ssn_stdp_noise}\crefanno{a1,a2}).
Consequently, wake-sleep learning is significantly improved with \gls{sal}, facilitating a more accurate learning of target distributions even when symmetric wake-sleep plasticity remains active (\cref{fig:results:ssn_stdp_noise}\crefanno{a1,a2}).

Looking at the results for \gls{kp}, we notice that weight decay actually hampers learning of good probabilistic representations in the presence of plasticity noise (see \cref{fig:results:ssn_stdp_noise}\crefanno{b2}) for all noise levels. %
Because weight decay is applied symmetrically to both weights, it is not capable of capturing the diverging pairs of weights due to asymmetric weight updates coming from the plasticity noise.
Therefore, the best performances (lowest \glspl{dkl}) are obtained for weight decay rates so small that \gls{kp} is practically disabled, as demonstrated in \cref{fig:sm:kp_lr}.
While having chosen this setup for its generality, we note that an equivalent effect occurs in any networks with a nonzero amount of neuronal diversity.
Because in such networks nominally identical weights would lead to different \gls{psp} sizes and shapes, identical weight updates for either functional or symmetrizing plasticity mechanisms do not amount to identical \emph{effective} weight updates, which is, however, the quantity that ultimately matters.

\subsubsection{Application 2: Cortical microcircuits}
\label{sec:results:mc}

\begin{figure*}[htb]
    \centering
    \begin{tikzpicture}
        \node[anchor=north west] at (-0.1, 0.0) {\includegraphics[width=3.0cm]{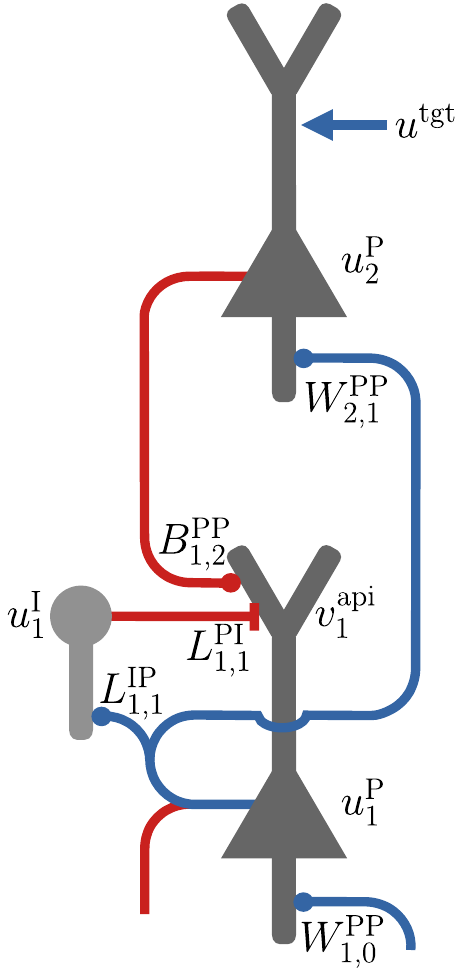}};
        \node[anchor=north west] at (0.0, 0.0) {\figanno{a}};
        \node[anchor=north west] at (3.3, 0.0) {\includegraphics[width=6.0cm]{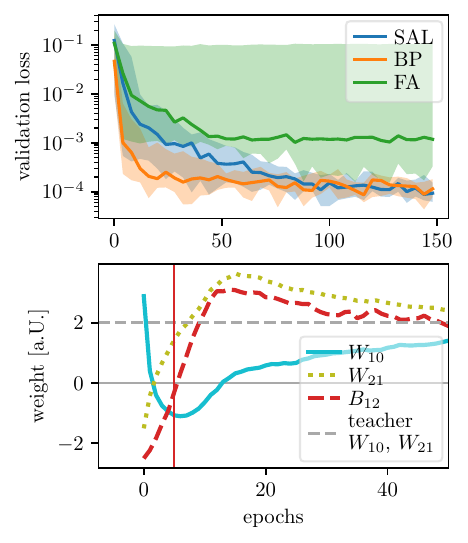}};
        \node[anchor=north west] at (3.2, 0.0) {\figanno{b1}};
        \node[anchor=north west] at (3.2, -3.5) {\figanno{b2}};
        \node[anchor=north west] at (9.9, 0.0) {\includegraphics[width=6.0cm]{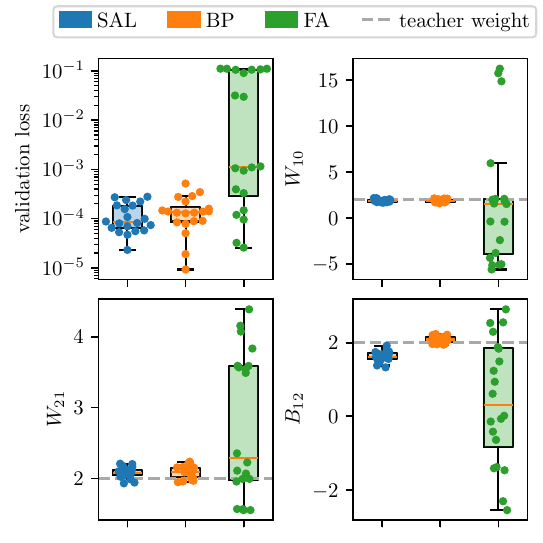}};
        \node[anchor=north west] at (9.7, 0.0) {\figanno{c}};
    \end{tikzpicture}
    \caption{\captiontitle{\gls{sal} enables accurate \gls{bp} in a spiking cortical microcircuit model.}
    \captionanno{a} Cortical microcircuit model for biologically plausible error backpropagation based on \cite{sacramento2018dendritic}, which we augment with a spiking mechanism.
    \captionanno{b1} Performance of different learning rules on a teacher mimic task.  \gls{sal} outperforms \gls{fa}, and training performance is on par with weight copying (\gls{bp}).
    Note the large performance variability of \gls{fa}, which depends critically on the initial weights.
    \captionanno{b2} Evolution of top-down and bottom-up weights during \gls{sal} learning for a case where initial feedback weights carry the wrong sign.
    Forward weights $\mWPP_{1,0}$  first evolve in the wrong direction (with \gls{fa}, they would explode, see h), but the gradual alignment of the backward weights $\mBPP_{1, 2}$ improves the feedback error signals, ultimately recovering the performance of vanilla BP.
    The red vertical line indicates the time point at which $\mBPP_{1, 2}$ switches sign.
    \captionanno{c} Distribution of final validation loss and learned weights in the student network.}
    \label{fig:results:mc}
\end{figure*}

We now consider a specific implementation of \gls{sal} in dendritic cortical microcircuits.
Such units have been suggested to enable the transmission of forward signals and backward errors across cortical hierarchies, realizing a biologically plausible variant of gradient descent through \gls{bp}~\cite{sacramento2018dendritic}.
To our knowledge, our implementation represents the first spiking version of this microcircuit model, thereby offering an alternative to other models of spike-based backpropagation such as burstprop~\cite{payeur2021burstdependent} or deep learning with segregated dendrites~\cite{guerguiev2017deep}.

Each microcircuit comprises two types of neuron populations: pyramidal cells and interneurons.
These neurons are organized into layers that correspond to cortical areas, following a biologically plausible connectivity pattern (see \cref{fig:results:mc}\crefanno{a}).

\paragraph{Student-teacher task}

In the following, we explore the model in a student-teacher task. 
Here, a chain of two pyramidal neurons is trained with \gls{sal} and, for comparison, with \gls{fa} and \gls{bp} as a reference.
A teacher configuration of the same size and depth is set up with fixed target weights to produce a non-linear input-output mapping.
The objective for the student circuits is to learn this function by adapting their weights accordingly.

A crucial aspect of this task is that successful learning requires the simultaneous adaptation of all the synapses in the network.
For the latent layer, it is essential that a meaningful error signal is transmitted through the feedback connections.
To illustrate the limitations of \gls{fa}, we conducted experiments where the feedback weights were randomly initialized;
i.e., in $\sim 50\,\%$ of experiments, feedback weights were initialized with opposite sign compared to feedforward weights.
Consequently, in these scenarios, the transported error also had the wrong sign, causing weight updates in the incorrect direction, and resulting in high average losses with a very large spread towards even higher values (\cref{fig:results:mc}\crefanno{b1}) and a drive of the weights away from their target (\cref{fig:results:mc}\crefanno{c}, green).

In contrast to the previous experiments with \glspl{ssn}, here \gls{sal} was only applied to the backward weights $\mBPP_{1, 2}$.
This allows $\mBPP_{1, 2}$ to quickly align with $\mWPP_{2, 1}$ at the start of training and subsequently follow its target, as shown in \cref{fig:results:mc}\crefanno{b1,b2}.
This facilitates fully local learning from any initial weight configuration: in \cref{fig:results:mc}\crefanno{b2}, $\mBPP_{1, 2}$ is initially set with the incorrect sign (indicated by the red dashed line), which should have been positive.
As a result, $\mWPP_{1,0}$ receives an erroneous error signal and decreases its value during the first 5 epochs.
With \sal, $\mBPP_{1, 2}$ learns to eventually switch to the correct sign.
From this moment on, a meaningful error signal is induced in $\vvApi_{1}$, enabling $\mWPP_{1, 0}$ to converge to its target value (indicated by the gray dashed line).

The study highlights the importance of correct error signal transmission through feedback connections for successful learning, demonstrating the limitations of \gls{fa} and the effectiveness of \gls{sal} in aligning backward connections with feedforward weights, thus ultimately facilitating fully local \bp learning.

\subsubsection{Application 3: Deep neural networks}
\label{sec:results:symmnet}

\paragraph{Symmetrization in two layers}

Before we turn to training a functional deep neural network with gradient descent to standard benchmark datasets, we conduct a preliminary experiment with only two layers similar to the design proposed in \cite{ahmad2020overcoming}, in which we compare the symmetrization capabilities of \gls{sal} against the two other spiking algorithms \gls{rdd}~\cite{guerguiev2017deep} and \gls{stdwi}~\cite{ahmad2020overcoming}.

The network consists of two densely connected layers with 50 neurons each and with randomly initialized synaptic weights.
The initial forward and backward weights $\mat W$ and $\mat B$ are independently drawn from a uniform distribution according to the Kaiming He initialization scheme~\cite{he2016deep}.
During symmetrization, forward weights are constant, and the different learning rules are applied only to the backward weights.
The interplay between symmetrization and functional credit assignment requires the backward weight updates to be fast enough to keep up with the changing forward weights.
For this reason, we hand-tune the learning rates such that the algorithms converge withing ca. \SI{1000}{\second}.
An in-depth discussion about convergence speeds follows at the end of this section.

\begin{figure}[htb]
    \centering
    
    \begin{tikzpicture}
        \node[anchor=north west] at (0.0, 0.0) {\includegraphics[width=0.95\linewidth]{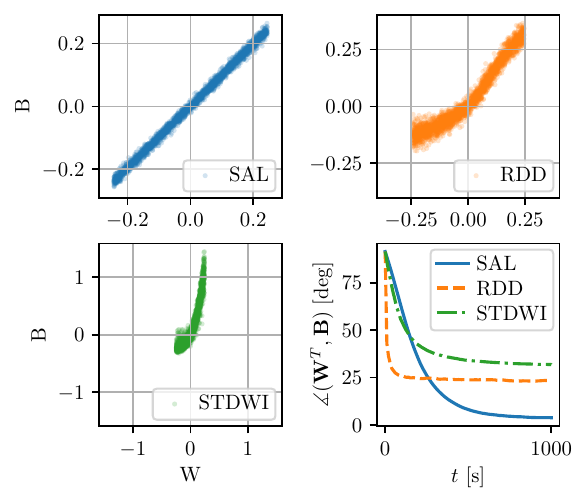}};
        \node[anchor=north west] at (0.3, -0.2) {\figanno{a}};
        \node[anchor=north west] at (4.2, -0.2) {\figanno{b}};
        \node[anchor=north west] at (0.3, -3.5) {\figanno{c}};
        \node[anchor=north west] at (4., -3.5) {\figanno{d}};
    \end{tikzpicture}
    \caption{
        \captiontitle{
            \gls{sal} performs accurate weight transport between densely connected layers.
        }
        Three spiking symmetrization algorithms \gls{sal}, \gls{rdd} and \gls{stdwi} are applied to align the backward weights with randomly drawn, fixed forward weights.
        \captionanno{a-c} The scatter plots depict the final value of the backward weights compared to their forward weight targets.
        \captionanno{d} Time evolution of the alignment angle for the three algorithms.
    }
    \label{fig:results:scatter}
\end{figure}

In \cref{fig:results:scatter}, we show scatter plots of the final backward weights compared to their target forward weights as well as the cosine similarities between $\mat W$ and $\mat B$ as a function of time during training.
While all algorithms succeed in aligning the sign between $\mat B$ and $\mat W$, \gls{sal} clearly stands out by carrying out the most precise weight transport.

In particular, \gls{stdwi} is particularly prone to deviations from the expected diagonal.
This behavior stems from a specific design choice in the original \gls{stdwi} network model, which uses \enquote{pseudo backward connections}, i.e., synapses that can be updated but have no effect on their post-synaptic neurons.
Consequently, the \gls{stdwi} weight updates uniquely depend on $\mat W$ and receive no feedback from potentially converging $\mat B$s, preventing the learning from stopping upon reaching the correct weight value.
Instead, \gls{stdwi} weight evolution stops only due to a weight decay term in the update rule.
Hence, the scale of the backward weights is rather set by the choice of hyperparameters.
Therefore, the deep network simulations below only include \gls{sal} and \gls{rdd} in the following.

\begin{figure*}[htb]
    \centering
    \begin{tikzpicture}
        \node[anchor=north west] at (-0.0, 0.0) {\includegraphics[width=3.7cm]{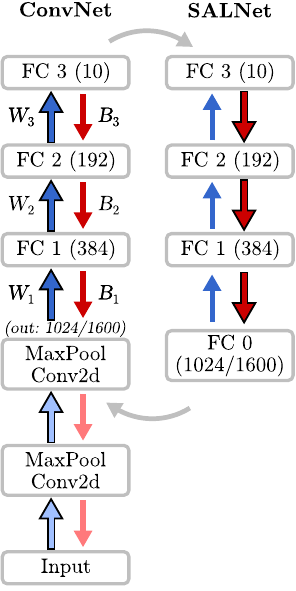}};
        \node[anchor=north west] at (4.0, 0.0) {\includegraphics[width=13.0cm]{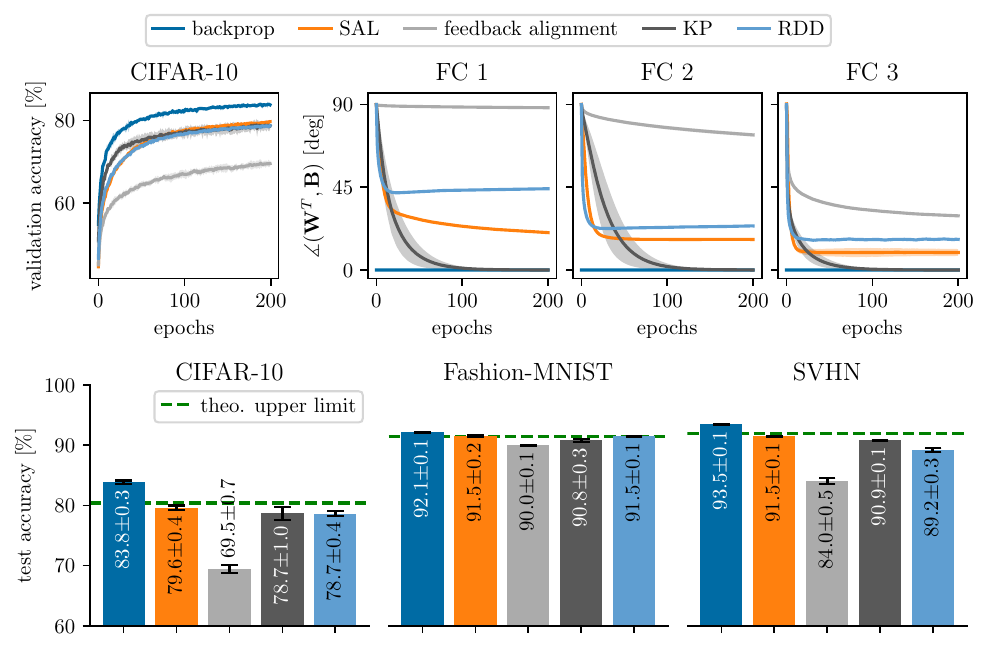}};
        \node[anchor=north west] at (-0.3, 0.0) {\figanno{a}};
        \node[anchor=north west] at (4.3, -0.5) {\figanno{b1}};
        \node[anchor=north west] at (8.0, -0.8) {\figanno{b2}};
        \node[anchor=north west] at (4.3, -4.3) {\figanno{c}};
    \end{tikzpicture}
    \caption{
        \captiontitle{Comparison of \gls{sal} with other symmetrization schemes in a deep network.}
        \captionanno{a} Architecture of the network.
        A convolutional neural network for learning the forward weight by gradient descent is paired with a \gls{snn} for learning the backward weights in a phased manner.
        The \gls{convnet} consists of two convolutional layers and three fully connected~(FC) layers; the FC are  interfaced with the spiking network.
        \captionanno{b1} Validation accuracy and \captionanno{b2} alignment angles in the FC layers for CIFAR-10 and the different symmetrization schemes.
        \captionanno{c} Comparison of the final validation accuracies for CIFAR-10, Fashion-MNIST and SVHN.
        We report the mean and standard deviation over five seeds in all panels.
        The green dashed line indicates the theorical upper limit for any symmetrization algorithm, i.e., a network that is trained with \gls{bp} in the FC layers and with \gls{fa} in the convolutional layers.
    }
    \label{fig:results:symmnet}
\end{figure*}

\paragraph{Symmetrization in deep networks}
To assess the efficacy of \gls{sal} in deep neural networks and compare it to other symmetrization schemes, we apply the benchmarking method used in \cite{guerguiev2019spikebased}.

In this method, learning of forward and backward weights is carried out in two separate networks:
1)~Forward weights $\mat W$ and biases $\vv b$ of a classical \gls{convnet} are trained via stochastic gradient descent to solve an image classification task with fixed feedback weights $\mat B$ during the forward learning phases.
2)~The weights $\mat B$ are learned in a separate \gls{snn} with spiking methods such as \gls{sal} and \gls{rdd} \cite{guerguiev2019spikebased} prior to each epoch.
Training of the full, two-part architecture proceeds in alternating fashion, transferring weights between the \gls{convnet} and \gls{snn} before each respective phase. %
This scheme isolates the symmetrization method from the functional plasticity algorithm, while also allowing a quantitative comparison of both spike- and rate-based approaches to learning the feedback weights.

Our two-part network architecture called \gls{symmnet} is laid out in \cref{fig:results:symmnet}\crefanno{a}.
The \gls{convnet} consists of two convolutional layers, followed by three \gls{fc} layers.
The \gls{snn} shares the architecture of the three \gls{fc} layers, meaning that the \gls{convnet}'s weights are only symmetrized between these layers.
We name the \gls{snn} by SNNs according to their respective symmetrization algorithms \gls{salnet} and \gls{rddnet}.

For comparison, we also train the \gls{convnet} without symmetrization via the \gls{snn}, using classical \gls{bp}, \gls{fa} and a weight decay algorithm (Kolen \& Pollack~\cite{kolen1994backpropagation}, Akrout et al.~\cite{akrout2019deep}).
We report the results for three different datasets: CIFAR-10 \cite{krizhevsky2009learning}, Fashion-MNIST \cite{xiao2017fashion} and SVHN \cite{netzer2011reading}.
The results are displayed in \cref{fig:results:symmnet}\crefanno{b1} and \crefanno{c}.

For all three datasets, \gls{bp} leads to the best validation accuracies, while \gls{fa} consistently yields the lowest accuracies.
The three symmetrization schemes reside between them, with \gls{sal} slightly outperforming the other two across all studied datasets.
While such similarity in performance might suggest a similarity in the trajectories of the learned weights, there are important differences between these algorithms, as we discuss below.

\Cref{fig:results:symmnet}\crefanno{b2}, we show the alignment angle between $\mat W_\ell$ and $\vv B_\ell$ (cosine similarity) for all three layers.
As expected, \gls{fa} results in the largest degrees of misalignment, explaining why \gls{fa} achieves the lowest accuracy values.
\gls{kp}, on the other hand, is able to perfectly symmetrize the weights after ca.~100 epochs.
However, this does not necessarily lead to the best accuracy of all symmetrization schemes\, as the weight decay term acts on the forward weights as an L2 regularizer. %
Hence, symmetrization and regularization are always coupled, such that any choice of the weight decay rate $\lambda$ is always a (potentially suboptimal) compromise between symmetrization and regularization. %
\Gls{sal} however acts only on the backward weights in this setup, without any direct interference with the forward learning.
In turn, \Gls{rdd} produces larger alignment angles than \gls{sal}, which manifests itself in slightly lower accuracy than \gls{sal}.

\paragraph{Time scale of symmetrization with \gls{sal}.}
\label{sec:results:salnet_timeevo}

\begin{figure}[htb]
    \centering
        \includegraphics[width=\linewidth]{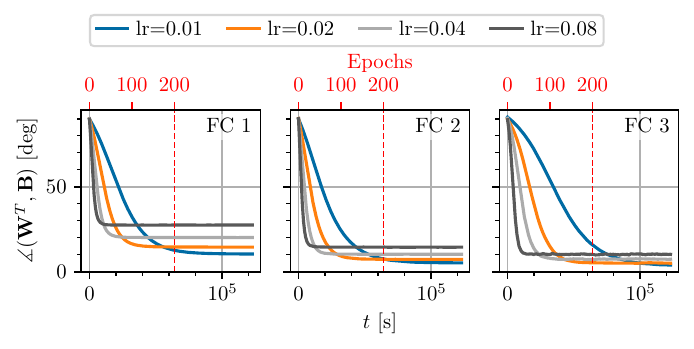}
    \caption{
        \captiontitle{Time evolution of \gls{sal} in the SALNet} for the three layers with constant forward weights for different learning rates.
        The learning rate serves as the crucial tuning parameter in the trade-off between fast symmetrization and small alignment angles.
        The upper x-axis displays the number of training epochs in \cref{fig:results:symmnet} corresponding to the total \gls{sal} sampling time across the entire training.
    }
    \label{fig:results:salnet_timeevo}
\end{figure}

In principle, no lower bound for the alignment angle exists when \gls{sal} is employed with infinitely small learning rates and arbitrarily long sampling time.
However, when learning to perform real-world tasks, symmetrization is only useful if it can be achieved quickly enough.
\Cref{fig:results:salnet_timeevo} shows the time evolution of the alignment angles for fixed $\mat W$ in the \gls{convnet}.

How %
does this compare with the time required for learning the forward weights?
Due to the lack of internal temporal dynamics, feedforward \glspl{ann} lack a natural sense of time.
Hence, for a physical (spiking) implementation of the ANN, we need to assign a physical processing time to a single input or an epoch.
As a reference, we estimate the fastest processing time scales of visual input in early visual cortex when considering purely feedforward processing of no more than one output spike per neuron (see \cite{goeltz2021fast,goeltz2025delgrad})
rather than rates, which would require longer integration times.
In such a fast processing scheme, each layer reacts on a time scale determined by its dominant time constant -- in this case, the membrane time constant $\tau \approx \SI{10}{\milli\second}$.
Accounting for transmission delays in the same order of magnitude, we estimate a total processing time of ca. $t \approx N \cdot 2 \tau$, where $N$ is the number of layers in the network.
For a visual hierarchy with 5 consecutive areas/layers (as in the ConvNets above), this yields a processing time of about $\SI{100}{\milli\second}$, in line with observations in the primate brain \cite{dicarlo2012how}.
At 50.000 images per epoch, this yields a total of 5000 s, an order of magnitude above the ensuing SAL phase of only 320 s per epoch.
Thus, even if SAL only occurs in phases rather than simultaneously with forward plasticity, the induced overhead appears negligible.
After a few epochs, as soon as the $\mat W$s only change gradually and the $B$s have already approached their targets, the \gls{sal} learning rate and the phase duration can be reduced to achieve the necessary small corrections for $\mat B$.

Note that when functional forward learning is activated, the $\mat W$-$\mat B$ alignment can be significantly accelerated, because the $\mat W$s also tend to align with the $\mat B$s in the first couple of epochs.

\subsection{Compatibility with Dale's law}
\label{sec:results:dales_law}

Until now, we neglected Dale's law: the empirical observation that a neuron can only express one type of neurotransmitter, hence being either excitatory or inhibitory.
In the following, we demonstrate that \gls{sal} works equally well in networks of neurons obeying Dale's law. %
As previously emphasized, what truly matters for the information exchange between neurons is the \emph{effective} weight $\Weff$, i.e., the totality of all influences of one neuron on another, through direct and indirect connections.
For example, an excitatory neuron can project directly onto another neuron, and indirectly via an inhibitory interneuron.  The effective weight therefore depends on the relative strength of the two paths,
also permitting sign flips as either of the two weights evolves during learning.

\begin{figure}[htbp]
    \centering
    \begin{tikzpicture}
        \node[anchor=north west] at (-0.0, -.0) {\includegraphics[width=0.35\linewidth]{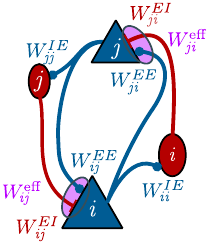}};
        \node[anchor=north west] at (0.4\linewidth, 0.0) {\includegraphics[width=0.6\linewidth]{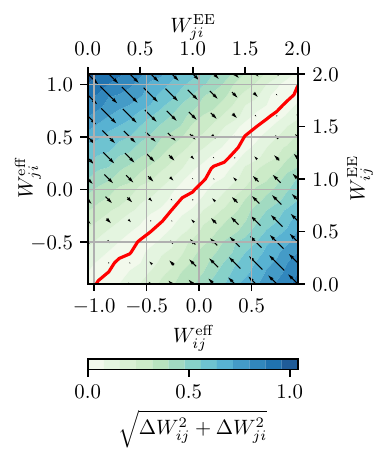}};
        \node[anchor=north west] at (0.0, -3.5) {\includegraphics[width=0.45\linewidth]{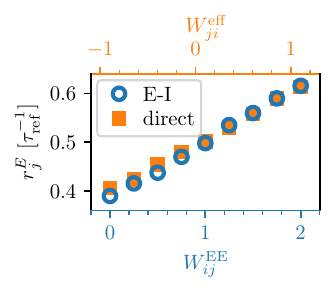}};
        \node[anchor=north west] at (-0.0, 0.0) {\figanno{a}};
        \node[anchor=north west] at (0.4\linewidth, 0.0) {\figanno{c}};
        \node[anchor=north west] at (0.0\linewidth, -3.7) {\figanno{b}};
    \end{tikzpicture}
    \caption{
        \captiontitle{Symmetrization and sign changes in networks obeying Dale's law.}
        \captionanno{a} E-I network: an augmented two-neuron system with (direct) excitatory and (indirect) inhibitory pathways.
        The effective weights $\Weff_{ij}$ are approximately equivalent to the sum of $W^{EE}_{ij}$ and $W^{EI}_{ij}$, allowing sign changes in $\Weff$, even if only $W^{EE}$ or $W^{EI}$ change.
        \captionanno{b} Mean firing rates of excitatory neuron 2 when \gls{sal} acts only on $W^{EE}_{21}$ compared to the equivalent 2-neuron circuit that neglects Dale's law.
        \captionanno{c} \Gls{ppd} of the E-I network, where \gls{sal} acts only on $W^{EE}_{ij}$. 
    }
    \label{fig:results:ei_eff_weights}
\end{figure}

As in \cref{sec:results:sal_spiketiming}, we can thus zoom onto a pair of neurons within a larger network and augment it with indirect inhibitory pathways in both directions (\cref{fig:results:ei_eff_weights}\crefanno{a}).
For simplicity, the inhibitory neurons obey the same stochastic dynamics as the excitatory neurons and receive strong input from their excitatory afferents, ensuring a correspondingly high correlation in their output activity.
Consequently, the \glspl{psp} from the direct and indirect pathways roughly coincide, with an effective weight of approximately $W^\mathrm{eff}_{ij} \approx W^{EE}_{ij} + W^{EI}_{ij}$ (\cref{fig:results:ei_eff_weights}\crefanno{b},  see also \cref{eq:methods:w_eff}).

\Gls{sal} uses the entire \gls{stdd} for symmetrization, which is influenced by both the direct excitatory and the indirect inhibitory pathway.
Therefore, \gls{sal} always symmetrizes towards the effective weight even when applied at only one of the synapses, and can cause sign changes in \Weff by altering the ratio of inhibition to excitation.
For example, SAL may only act on synapses between excitatory neurons $W^{EE}$.
In \cref{fig:results:ei_eff_weights}\crefanno{c}, we show a \gls{ppd} of this system, demonstrating that \gls{sal} is able to reliably symmetrize weights while obeying Dale's law.

\subsection{Other PSP shapes and their effect}
\label{sec:results:other_psps}

\begin{figure*}[htb]
    \centering
    \begin{tikzpicture}
        \node[anchor=north west] at (0.0, 0.0) {\includegraphics[width=0.9\linewidth]{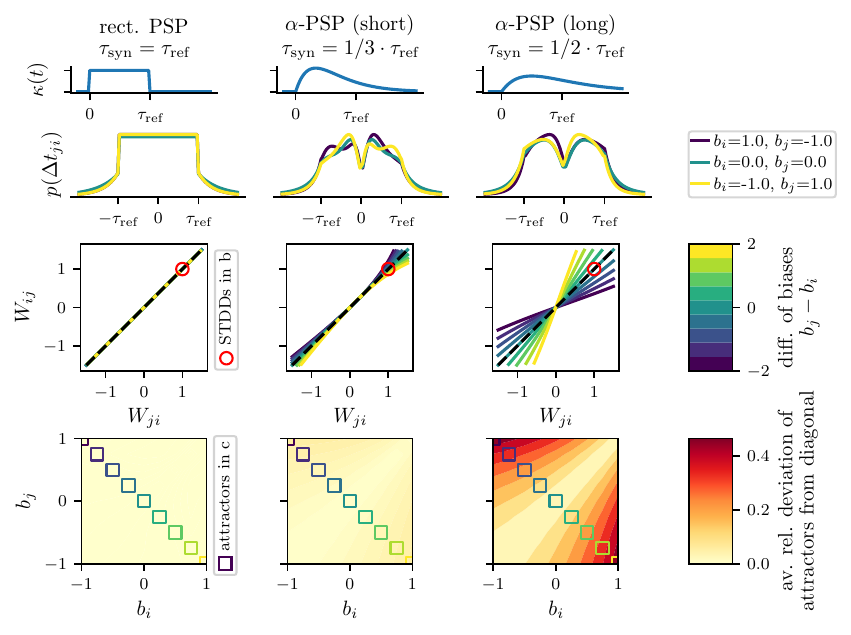}};
        \node[anchor=north west] at (0.3, -0.5) {\figanno{a}};
        \node[anchor=north west] at (0.3, -1.9) {\figanno{b}};
        \node[anchor=north west] at (0.3, -4.0) {\figanno{c}};
        \node[anchor=north west] at (0.3, -7.0) {\figanno{d}};
        
    \end{tikzpicture}
    \caption{
        \captiontitle{\Gls{sal} works with different \gls{psp} shapes}.
        \captionanno{a} We consider a two-neuron system together with three common \gls{psp} kernels. 
        \captionanno{b} \Glspl{stdd} at $W_{12} = W_{21} = 1$ for three different bias combinations. The \glspl{stdd} of the $\alpha-$kernels show a slight asymmetry if $b_1 \ne b_2$.
        \captionanno{c} Phase plane diagram of reciprocal weights for nine bias combinations.
        For clarity, only the attractors are shown. 
        The biases used here are marked as squares in the respective color in panel~d).
        Rectangular \glspl{psp} always yield perfect symmetrization (all attractors are located on the diagonal).
        On the other hand, symmetrization is not perfect for $\alpha-$\glspl{psp}.
        The attractor deviates from the diagonal if the difference in biases is large, with longer \glspl{psp} causing larger deviations.
        Importantly, however, the sign of the weights is always aligned correctly.
        Red circles indicate the weights used for the \glspl{stdd} above.
        \captionanno{d} Average relative deviation of the attractor from the diagonal as a function of biases $b_1$ and $b_2$.
        Rectangular \glspl{psp} symmetrize perfectly for all biases.
        Short \glspl{psp} (middle) only cause small deviations (less than \qty{10}{\percent} across the whole bias-space), while long \gls{psp} tails incur larger deviations~(right).
    }
    \label{fig:results:psps}
\end{figure*}

So far, we have only considered rectangular \gls{psp} shapes.
More biologically plausible \glspl{psp} exhibit a more complex behavior when using \gls{sal}, effectively introducing a bias-dependent behavior in the symmetrization (\cref{fig:results:psps}).
The bias serves as a proxy for the base firing rate, which can result from neuron-specific leak potentials or different input rates from the surrounding networks, averaged over time.
In the following, we discuss how \gls{sal} behaves in complex networks when other \glspl{psp} shapes are used.

We again zoom onto a pair of neurons within a larger network.
As the most important feature for evaluating the effectiveness of \gls{sal}, we assess the shape of the basin of attraction in the $W_{ij}$/$W_{ji}$ plane as visualized by the \gls{ppd} in \cref{fig:results:sal}\crefanno{e}.
The basin of attraction provides information about the weight pairs to which \gls{sal} converges.

We employ an \ensuremath{\alpha}-shaped kernel 
\begin{equation}\label{eq:results:alpha_psp}
  \psp(t) = \Theta(t) \frac{\tauref}{\tausyn^2}\, t \exp\left( -\frac{t}{\tausyn} \right),
\end{equation}
with different synaptic constants $\tausyn$ (see \cref{fig:results:psps}\crefanno{a}) to model typical \gls{psp} shapes found in both biological and neuromorphic systems.
In \cref{fig:results:psps}\crefanno{c}, we visualize the attractors of the rectangular and $\alpha$-shaped \glspl{psp} with a short and long time constant $\tausyn$.
For the sake of clarity, we do not plot the flow field, but concentrate on the shape of the attractors.
Notably, \gls{sal} converges in all cases, i.e., all fixed points are indeed stable and remain organized as line attractors, ensuring that \gls{sal} gives rise to useful weight updates.
For a formal proof, we refer to \cref{sec:methods:proof}.
Furthermore, the signs of the weights are matched in all cases (i.e., $\sgn(W_{ij}) = \sgn(W_{ji}) $), independently of the biases and the \gls{psp} shape.
This is an important result, as
sign conservation has been shown to greatly reduce the gap between \gls{fa} and \gls{bp} \cite{moskovitz2018feedback}.

These deviations from perfect weight mirroring are best understood by considering the effect of \gls{psp} shapes on the \stdd (\cref{fig:results:psps}\crefanno{b}).
Here, we provide a qualitative explanation, but refer to \cref{sec:methods:analytic_stdd} for an exact analytical result.
Since the firing probability is a monotonic function of the membrane potential, we expect the shape of the \stdd to roughly follow that of the \glspl{psp}.
Indeed, for rectangular \glspl{psp}, we observe a flat distribution, as a constant \gls{psp} leads to a uniform spiking probability in the postsynaptic neuron $j$ for all $\DeltaT < \tauref$.
On the other hand, for $|\DeltaT| > \tauref$, the spike timing differences are produced by two independent Poisson processes, giving rise to a left-right symmetric exponential distribution.

For $\alpha$-\glspl{psp}, the \stdd shape is still similar to that of the \gls{psp}, but the strict left-right symmetry is broken.
We identify two main causes for this effect.

First, we note that $\alpha$-\glspl{psp} extend beyond the refractory period and can therefore stack, unlike the rectangular \glspl{psp} with $\tausyn=\tauref$.
The probability of such stacking increases with the presynaptic neuron's firing rate, which, in turn, is determined by its bias.
Unequal biases lead to asymmetric stacking probabilities and therefore to asymmetric \stdds.
Furthermore, this stacking also means that the spiking processes are not independent anymore for $|\DeltaT| > \tauref$.
However, this phenomenon has a much smaller effect due to the exponential decay of the \sal \stdp kernel.
Evidently, these stacking effects are altogether more pronounced for longer \glspl{psp}, and thus represent the main reason for the discrepancies in \cref{fig:results:psps}\crefanno{b-d}, right panels.

The second cause of \stdd symmetry breaking is best observed in \cref{fig:results:psps}\crefanno{b}, middle panel, where a second bump around $\pm \tauref$ stands out.
Consider the yellow curve, where neuron $j$ has a high bias and neuron $i$ a low one.
If neuron $i$ fires, then neuron $j$ is more likely to fire in quick succession ($i \to j$), leading to the small causal \gls{stdd} peak for small positive values of $\Delta t_{ji}$.
In turn, if neuron $j$ fires, then neuron $i$ is more likely to fire in response ($j \to i$), leading to the anticausal peak for small negative values of $\Delta t_{ji}$; this peak is larger because this order of spiking happens more often, as neuron $j$ is overall more likely to fire due to its higher bias.
However, also due to this high bias, it is more likely to fire again on its own as soon as its refractory period ends ($j \to i \to j$).
This creates the third peak around $\Delta t_{ji} \lesssim \tauref$, which appears on the causal branch but is merely an echo of the anticausal peak caused by the initial firing of neuron $j$.
Since this effect is bias-dependent, it breaks the \stdd symmetry for unequal neuronal biases.
Being most pronounced for short, peaked \glspl{psp}, it represents the main cause behind the deviations in \cref{fig:results:psps}\crefanno{b-d}, middle panels.

We quantify the deviation of the attractor from the diagonal by computing the average relative distance to the diagonal as a function of $b_1$ and $b_2$ (see also \cref{sec:methods:psps}).
In general, the relative deviation is smaller for shorter \glspl{psp}, as well as for smaller differences in bias (\cref{fig:results:psps}\crefanno{d}, middle and right panel).
As explored in the previous section, such small amounts of asymmetry can be well tolerated in deep learning setups.
However, these bias-induced distortions can also be easily suppressed by introducing a dedicated \gls{sal} phase into the learning schedule, during which neuronal biases are reduced across the network.

\section{Discussion}
\label{sec:discussion}

Plausible algorithms for physical computing -- whether instantiated in neuromorphic hardware or biological brains -- need to tackle the weight transport problem.
In this paper, we have presented \gls{sal}, a fully local algorithm for effective weight transport in spiking neuronal networks based on \gls{stdp}.

\gls{sal} enables the alignment of bottom-up and top-down information flow (for instance, of the sensory information and error signals, depending on the computational paradigm \cite{huang2011predictive,millidge2021predictive,mikulasch2023error}),
thereby fulfilling the core requirement of credit assignment in physical systems: %
the locality of information in space and time. 
In the domain of neuromorphic computing, \gls{sal} can play a key role in the implementation of fully local on-chip learning schemes.
In models for computation in the brain, \gls{sal} can augment their biological plausibility when networks are subject to symmetry constraints.
Additionally, we emphasize \gls{sal}'s ability to increase the robustness of networks to the omnipresent parameter diversity found in any physical substrate.
More concretely, \gls{sal} automatically balances effective weights and does not just copy the numerical value. 
Thus, the effect of morphological variance of different neurons on computation is also compensated.
While such effects can also be accounted for by more complex learning rules based on non-Euclidean gradient descent \cite{kreutzer2022naturalgradient,surace2020choice}, \gls{sal} lends itself as a diversity-independent amendment to any functional learning rule, unlike other proposed symmetrization algorithms, as discussed below.

In contrast to other approaches, \gls{sal} is free of any (oftentimes implicit) assumption regarding the symmetry of shared parameters between the involved neurons and synapses:
Instead, \gls{sal} works independently in each synapse and compensates for asymmetric weight updates.
Moreover, \gls{sal} uses the omnipresent temporal noise in physical substrates as a resource: While noise in neural network is often regarded as destabilizing for computation, it actually drives the stabilizing effect of \gls{sal}.
Finally, we have also shown that \gls{sal} is mostly agnostic to the underlying network topology.
It is compatible with a plethora of network architectures that involve reciprocal connectivity.
In this context, we also highlight that it has allowed us to demonstrate the first functional spiking implementation of the backpropagating microcircuits proposed in~\cite{sacramento2018dendritic}.

\paragraph{Related work}
Broadly, existing approaches to address the weight transport problem can be divided into four categories:

First, some studies simply assume weight symmetry in their theoretical derivations and implement physically implausible copy operations in their simulations. 
While this allows to show that the respective algorithms are in general capable of successfully performing credit assignment, the absence of a solution to the weight transport problem calls their biological plausibility into question, while also limiting their applicability for neuromorphic on-chip learning.
An example here is equilibrium propagation \cite{scellier2017equilibrium} or the work by Xie and Seung~\cite{xie2003equivalence}.

Second, random, fixed backprojections as in~\gls{fa}~\cite{lillicrap2016random,nokland2016direct} are often cited as a one-fits-all approach to solve the weight transport problem.
However, numerous studies have highlighted its limitations when it comes to scaling to deeper networks or robustness \cite{bartunov2018assessing,moskovitz2018feedback,max2024learning}.
The detrimental effect of misaligned weights carries over to spiking networks, as we have shown here.
Importantly, it is known that weight alignment does not need to be perfect for transportation of meaningful gradients in deep networks~\cite{liao2016important,moskovitz2018feedback,xiao2018biologically,max2024learning,}. %
Instead, the conservation of the sign is the most important prerequisite for good performance, and SAL enables a degree of alignment that far exceeds this minimalist condition across a wide range of commonly used \gls{psp} shapes(see \cref{tab:sm:symmnet_extented}).

The third category is formed by algorithms that are based on weight decay.
The idea was first introduced by Kolen and Pollack \cite{kolen1994backpropagation}, using symmetric weight updates in the forward- and backward path in conjunction with a constant weight decay that causes the network to gradually forget its asymmetric initial state.~%
\cite{akrout2019deep} presents a neuronal circuit model that could instantiate such symmetric weight updates.
However, symmetrization then relies on strictly symmetric weight updates, and it is not clear how these symmetric updates can come about given heterogeneous synapses and plasticity (see \cref{fig:results:ssn_stdp_noise});
in that sense, weight decay algorithms only relegate the weight transport problem to a weight \emph{update} transport problem. %
By contrast, \gls{sal} relaxes all implausible constraints on symmetric updates because it is able to dynamically realign weights during training.

In the fourth category, we summarize algorithms that dynamically learn useful reciprocal weights in a physically plausible manner.
With (difference) target propagation, \cite{bengio2014how, lee2015difference, meulemans2020theoretical, meulemans2021credit} learn pseudo-inverse backprojections to perform Gauss-Newton optimization.
For error backpropagation, \cite{akrout2019deep, ernoult2022scaling, lansdell2019learning, max2024learning}
learn feedback weights that approximately align with the forward Jacobian, but are data-specific and do not necessarily match the forward weights in angle or amplitude.

Akrout et al. %
introduced in \cite{akrout2019deep} a second symmetrization algorithm called \gls{wm}, which is related to \gls{sal} in that it uses noise to enable weight transport, but \gls{sal}’s key advantage is that it operates directly in spiking networks and does not rely on linear activations for exact transport. 
In addition, \gls{wm} assumes purely uncorrelated noise and a specialized mirroring circuit with transient strong lateral connections and ideally sequential layer activations to avoid shared-input correlations, whereas SAL can tolerate correlated activity as long as each neuron receives enough uncorrelated noise, while also not requiring any additional circuitry.

A particular example from this class of algorithms is \gls{pal}~\cite{max2024learning}, as it can be regarded as the rate-based complement to \gls{sal}.
Notably, their common denominator is the functional role of noise:
In \gls{pal}, the information carrying signal is augmented by high frequency noise, which moves in loops through the network. 
On its way, the noisy signal is affected by the synaptic weights and therefore carries information that can be exploited locally to align weights.
A further commonality between \gls{sal} and \gls{pal} is their ability to train all backprojections in a network at the same time and without disturbing forward weight learning.

In the group %
of spiking algorithms for weight symmetrization, \gls{rdd}~\cite{guerguiev2019spikebased} and \gls{stdwi}~\cite{ahmad2020overcoming} are particularly relevant. 
Both algorithms rely on specific stimulation protocols, by which only a subset of neurons are stimulated at a time to sidestep cross-correlations between neuronal activity. 
This introduces the necessity of distinct symmetrization phases.
Also, it remains unclear how biologically plausible such tightly orchestrated stimulation protocols are and what their underlying neuronal mechanisms could be.

\Gls{rdd} %
uses the discontinuity at the spike threshold to infer the causal effect of spikes emitted by a neuron on the activity of downstream neurons and obtain an estimate for the optimal feedback weights.
This requires to store additional variables such as the \enquote{input drive} or the \enquote{post-synaptic spike responses} in each neuron and to fit a piece-wise (linear) model with four extra parameters in each synapse.
While this approach might be faster than \gls{sal} in simulated time (\SI{90}{\second} vs.~\SI{320}{\second}, see \cref{tab:methods:symmnet}), \cite{guerguiev2019spikebased} neither discusses the physiological properties in neurons that could give rise to these variables nor does it make statements about the implementability in current neuromorphic hardware.

More similar to \gls{sal}, \gls{stdwi} uses a three-factor \gls{stdp} rule with fast and slow eligibility traces to infer the forward weight and update the backward weight accordingly. 
However, this ansatz assumes \enquote{pseudo} backward synapses, i.e., information transporting connection whitout any influence on the postsynaptic membrane potential, an assumption which is difficult to reconcile with the  experimental reality of physical neuronal networks, biological and neuromorphic alike. 
Also, the question whether this algorithm scales in deeper networks and how well it integrates into algorithms for functional credit assignment remains open.

In \cite{burbank2015mirrored}, a spiking model for autoencoders using \gls{stdp} was introduced.
Similarly to \gls{sal}, an anti-Hebbian window (called \enquote{mirrored \gls{stdp}}) is employed to learn the backward projections of the autoencoder.
Despite the apparent similarity between \gls{sal} and mirrored \gls{stdp}, the two plasticity rules actually serve a different purpose:
In~\cite{burbank2015mirrored}, the combination of Hebbian and anti-Hebbian learning produces the same weight updates in the forward and backward direction.
As a result, mirrored \gls{stdp} alone cannot equilibrate initial weight differences.
Furthermore, it is not discussed whether the model can compensate for parameter noise on neuronal and plasticity parameters, which is an essential feature of \gls{sal} and decisive for its efficacy in neuronal systems with analog components, whether biological or neuromorphic.

A different approach to spike-based credit assignment is proposed in~\cite{oconnor2019training}.
Following the energy-based paradigm of equilibrium propagation, O’Connor et al.~propose a spiking stochastic approximation thereof.
While the authors demonstrate the feasibility of their approach, they also acknowledge that it lacks a solution to the weight transport problem;
this is exactly the gap which \gls{sal} is able to fill, and a combination of \gls{sal} with Spiking Equilibrium Propagation could shed light on neuronal credit assignment and drive efficient hardware implementations.

BurstProp \cite{payeur2021burstdependent}, a solution to credit assignment in layered spiking networks, uses a local plasticity rule together with postsynaptic bursts to determine the sign of the synaptic change of the forward weights.
However, the authors implement \gls{fa} for the transportation of the top-down teaching signal; as above, BurstProp may be combined with \gls{sal} to improve biological plausibility and learning performance.

The model of segregated dendrites~\cite{guerguiev2017deep} uses a compartmentalized neuron model in conjunction with phases to disentangle error and feedforward information. %
The error signal is fed back to each layer via fixed random feedback matrices, making it related to deep feedback alignment \cite{nokland2016direct}.
Since the information is encoded in the mean firing rate of Poisson point-process neurons, noise is inherently present so that \gls{sal} can be applied directly to better align forward- and backward pathways.

While%
not explicitly designed for weight symmetrization, we note a conceptual similarity between \gls{ng} learning (originally proposed in \cite{amari1998natural}, see \cite{kreutzer2022naturalgradient} for a spiking implementation) and \gls{sal}.
In \gls{ng} learning, weight updates do not follow the gradient of a cost function with respect to the synaptic weight, but rather with respect to the neuronal output.
This is similar to the way \gls{sal} operates on output spike times alone, leading to whatever weight change is necessary to shift the cost -- in this case, the asymmetry -- towards its desired value.

\paragraph{Applications and constraints}

Gradient descent algorithms, %
and in particular error backpropagation, represent some of the most powerful learning paradigms known to date.
This constitutes the primary reason for the considerable interest in their physical implementation, both in neurobiology and in neuromorphics (see, for example, \cite{richards2019deep,pfeiffer2018deep}).
All of these models require some form of weight transport, and attempts to sidestep this element (such as \gls{fa} or \gls{dtp})
lack both performance and scalability \cite{bartunov2018assessing}.
We thus believe that \gls{sal} closes a considerable gap in such models by proposing a concrete mechanism to either be experimentally validated (in biology) or implemented (in hardware).
While efforts in the latter category are currently underway, numerous proposals for error representation and transport in cortex have been put forward, many of them implicitly requiring a mechanism such as \gls{sal} for scalablility and performance \cite{whittington2017approximation, guerguiev2017deep, sacramento2018dendritic, payeur2021burstdependent, mikulasch2023error, max2025backpropagation}.
Because existing experimental data is not yet at the level of resolution required to (in)validate these hypotheses, including the required symmetrization mechanisms such as \gls{sal}, recent experimental collaborations have started gathering data for clarifying the microscopic underpinnings of such models \cite{aizenbud2025neural, microns2025functional}.

Aside from computation in the brain, \gls{sal} specifically targets spiking neuromorphic systems, especially if they contain analog components.
Such systems are prone to substrate variability and parameter drift, which destabilize training performance and limit real-world applicability.
This applies, for example, to emulations of quantum systems on analog spiking hardware~\cite{klassert2022variational,czischek2022spiking} or Bayesian inference using spiking nanolasers~\cite{boikov2025ultrafast}.
In all of these cases, asymmetries due to substrate variability constitute the main limiting factor for the task performance of the respective networks.

Importantly however, %
\gls{stdp}-like learning forms the standard for on-board/on-chip learning on many platforms~\cite{billaudelle2020versatile,schemmel2010wafer,moradi2018scalable,richter2024dynapse2,neckar2018braindrop,serrano2013stdp,boybat2018neuromorphic}. 
Therefore, \gls{sal} may provide the missing link for stable, physical learning systems:
by facilitating a fully local implementation of effective weight sharing, it can compensate parameter noise and drift.
Additionally, it paves the way for a fully local, analog implementation of \gls{bp}.

This capability positions \gls{sal} as a robust solution for adaptive systems, such as smart sensors and wearable devices, which must cope with challenges like sensor degradation or environmental changes while maintaining efficiency and autonomy.
In this context, we expect that memristive devices~\cite{serrano2013stdp,boybat2018neuromorphic}, known for their high density and low power consumption, provide an ideal substrate for implementing \gls{sal}'s learning rules.

While \gls{sal} %
can be applied during a dedicated symmetrization phase, this is often not necessary.
Instead, \gls{sal} can operate concomitantly with other (functional) learning rules, as explicitly shown in \cref{sec:results:ssn}, where \gls{sal} co-occurs with the sleep phase.
Similarly, as \gls{sal} can operate on only one of the reciprocal weights between a pair of neurons, for example the backward weights in functionally feedforward networks (\cref{sec:results:symmnet}), or the EE weights in networks that obey Dale's law (\cref{sec:results:dales_law}), it can be straightforwardly used in parallel with a different functional learning rule acting on the other one of the two reciprocal weights.

\Gls{sal} %
requires sufficient noise to enable synaptic weight measurements through the \gls{stdd}. 
Without adequate noise, weak or inhibitory synapses may fail to measurably influence postsynaptic firing, preventing proper sampling of spike-timing relationships.
While theoretically any noise level suffices under Gaussian assumptions (as firing probabilities remain nonzero), insufficient noise in practice can substantially slow convergence by limiting sampling frequency.
A practical heuristic is to set noise levels such that neurons fire at approximately half their maximum rate without synaptic input, though lower noise levels often suffice.

Noise can originate intrinsically in analog (including biological) substrates, be injected digitally via lightweight pseudorandom generators, or be provided as diffuse background spiking input (often modeled as Poissonian). Crucially, this noise need not be private per neuron: small deterministic networks or other network ensembles can supply shared pseudo-noise, and recurrent networks can even self-generate activity that is effectively stochastic \cite{dold2019stochasticity}.

Complementary to this, \gls{sal} is designed to address weight transport in spiking theories of the brain.
It suggests a functional need for diverse STDP curves, thus providing an explanation for their observation in nature~\cite{abbott2000synaptic,andrade2023timing}.

However, contrary to many hardware platforms \cite{billaudelle2020versatile,moradi2018scalable,richter2024dynapse2}, biological synapses cannot simply switch their sign, as they typically express only one type of neurotransmitter.
This makes biological synapses either excitatory or inhibitory, a phenomenon known as Dale's law~\cite{dale1935pharmacology}.
Many computational models ignore this detail by allowing a neuron to have both inhibitory and excitatory efferent synapses and also for a synapse to change its sign during learning.
While the first type of restriction can be lifted by appropriately including inhibitory neurons in all communication pathways, the sign-change problem requires more attention.
In \cref{sec:results:dales_law}, we have thus demonstrated the functionality of \gls{sal} in such a setup.
This demonstrates that  the effective weight -- the entirety of all connections, including direct and indirect pathways, multapses, etc. -- is the adequate equivalent to a weight in a deep learning model.
While we do not argue that this circuitry is an exact implementation of the brain's solution to the sign-changing problem, we provide a blueprint for further experimental investigations.
    
In particular, %
an essential feature of \gls{sal} in this context is its capability of symmetrizing \emph{effective} connection weights -- i.e., including the effect of inhibitory interneurons, multapses or other direct or indirect connection pathways -- even when applied to only a specific subset of all synapses.
Thus, unlike other rules such as \gls{kp}, which need to manifest in all synapses within a network, \gls{sal} would permit a greater diversity of synapse-specific plasticity mechanisms, thereby not only allowing greater freedom in the design of artificial spiking networks, but also better accommodating current experimental observations.

In summary, we contend that symmetry is not just a mere token of theoretical elegance, but also an important ingredient for practical deployment across a wide range of models and applications.
Instead of working around it when faced with its absence, \gls{sal} offers a solution for recovering and maintaining symmetry in real physical systems, by making use of noise as a fundamental computational ingredient.

\section{Methods} \label{sec:methods}
\resetlinenumber

\glsresetall  %

As in the main text, bold lowercase variables $\vv{x}$ denote vectors, bold uppercase letters $\mat{X}$ matrices.
We denote spike times of neuron $i$ as $\tspk_i$.
The most recent spike emitted by neuron $j$ before the time $t$ is denoted as~$\tspk'_i$.
The set of all past spikes from neuron $i$ is denoted by $\spkset{\tspk_i}$.
The output spike train from neuron $i$ is 
\begin{equation}\label{eq:methods:spiketrain}
    \spktrain_i(t) = \sum_{\spkset{\tspk_i}}{\delta(t - \tspk_i}),
\end{equation}
where $\delta(x)$ is the Dirac delta distribution.

\subsection{Neuron model}
\label{sec:methods:neuronmodel}

We use a \gls{glm} \cite{gerstner2014neuronal} to model the dynamics of spiking neurons.
The membrane potential of neuron $k$ is given by
\begin{equation}\label{eq:methods:mempot_gls}
    u_i(t) = b_i + \sum_{k} W_{ik}(t) \int \limits_{0}^{\infty} \psp(s) \spktrain_k(t - s) \dd s,
\end{equation}
where $b_i$ is the neuron's bias, which can be associated with the leak potential, $\psp(t)$ the kernel of the \gls{psp}, $\spktrain_k(t)$ the spike trains coming from the presynaptic neuron $k$, and $W_{ik}$ the corresponding synaptic weight.
If not stated otherwise, we employ a rectangular \gls{psp} kernel,
\begin{equation}\label{eq:methods:rect_psp}
    \psp(t)=\Theta(t) - \Theta(t - \tausyn) \,, 
\end{equation}
where $\tausyn$ denotes the synaptic time constant.

The output spikes of a neuron are generated by an inhomogeneous Poisson process with absolute refractory period of length $\tauref$.
The spiking probability in the time interval $[t; t + \dd t]$  is given by
\begin{equation}\label{eq:methods:probspk}
    p(\tspk_i| \tspk_i') = \begin{cases} r_i(t) \, \dd t & \text{if} \ t - \tspk_i' > \tau_\mathrm{ref} \\ 0 & \text{else,} \end{cases}
\end{equation}
where 
\begin{equation}\label{eq:methods:act_func}
    r_i(t) = \tauref^{-1} \exp \left( u_i(t) \right)
\end{equation}
is the instantaneous firing rate, rescaled by the refractory period.
The timescale of the rectangular \gls{psp} kernel is matched with the length of the refractory period $\tausyn = \tauref$.

We choose this model to account for stochastic firing pattern found throughout networks in the brain and to model the presence of temporal noise in the nervous system while being mathematically tractable.
It can be shown that \gls{lif} neurons in the high-conductance driven by high-frequency spike noise state can have similar statistical properties as the model presented above \cite{petrovici2016stochastic}.

\subsection{Spike-based alignment learning}
\label{sec:methods:sal}

The working principle of \gls{sal} builds on the observation that weights leave a characteristic imprint on the distribution of spike timing differences of pre- and postsynaptic spikes.
In \cref{sec:results:sal_definition} and \cref{fig:results:sal,fig:results:psps}, we have given a qualitative understanding of our weight alignment mechanism, which we now follow up with a more detailed explanation.

Typically, unequal reciprocal weights $W_{ij} \ne W_{ji}$ produce skewed/asymmetric \glspl{stdd}.
This is informative of the weight difference and can be exploited for symmetrization.
This asymmetry can be extracted by a standard \gls{stdp} rule, which updates $W_{ij}$ after every occurrence of a nearest-neighbor spike pair with $\DeltaT_{ij} = \tspk'_i - \tspk'_j$ via
\begin{equation}\label{eq:methods:stdp}
    \Delta W_{ij} = \eta \Big( \underbrace{\Theta(-\DeltaT_{ij}) \, \stdpAntiCausal \stdpwindowAntiCausal(\DeltaT_{ij})}_\text{anti-causal} + \underbrace{\Theta(\DeltaT_{ij}) \, \stdpCausal \stdpwindowCausal(\DeltaT_{ij})}_\text{causal} \Big).
\end{equation}
Here, $\stdpwindow(\DeltaT)$ defines the shape of the learning window, which we choose to be $\stdpwindow(\DeltaT) = \exp(t / \tau)$ in accordance with \cite{bi1998synaptic}. 
$\stdpCausal$ ($\stdpAntiCausal$) is the prefactor for the (anti)causal branch, which we set to be +1 $(-1)$ to strengthen (weaken) $W_{ij}$ if a (anti)causal spike pair occurs.

In practice, \gls{sal} works with spike pairs of infinite-range as long as the number of causal and anti-causal $\DeltaT$ are the same; i.e.~on average, a spike contributes as many times to causal $\DeltaT$ as to anti-causal ones;
if this is not the case, weights do not necessarily converge to a fixed point when using \gls{sal}.
In fact, the experiments in \cref{sec:results:experiments} use correlations of \emph{all} spike pairs, demonstrating that \gls{sal} also works beyond the strict analytical regime of nearest neighbor spike pairs.
The infinite-range \gls{stdp} rule can be written as
\begin{align}\label{eq:methods:stdp_2}
    \dot{W}_{ij}(t) = \eta \, \big(& \spktrain_j(t) \int \limits_0^\infty \stdpAntiCausal(s) \stdpwindowAntiCausal(s) \spktrain_i(t-s) \dd s \nonumber \\
    &+ \spktrain_i(t) \, \int \limits_0^\infty \stdpCausal(s) \stdpwindowCausal(s) \spktrain_j(t-s) \dd s \big)
\end{align}

\subsubsection{Analytical calculation of the spike-timing difference distribution}
\label{sec:methods:analytic_stdd}

Our analytical computation of the \gls{stdd} builds on the insight that a time discretized version of the \gls{glm} can be expressed as a discrete-time Markov chain.
The following derivation is inspired by \cite{buesing2011neural}.
We introduce a discrete counting variable $\zeta_i\in\setNat$, that defines the state of neuron~$i$ and a state vector $\vv{\zeta}\in\setNat^N$ collecting the states of all~$N$~neurons in the network.
Due to the Markov property, the network state $\vv{\zeta}'$ of the next time step depends only on the current state $\vv{\zeta}$.%

$\zeta_i$ counts the number of time steps since neuron $i$ has spiked the last time.
Depending on its own state $\zeta_i$ and the states of the other neurons $\zeta_k$, neuron $i$ can spike with some probability, for which we set $\zeta_i' = 1$.
If it doesn't spike, the counter is incremented, $\zeta_i' = \zeta_i + 1$.

For each neuron, we define analogous to \cref{eq:methods:mempot_gls} a membrane potential
\begin{equation}
    u_i = b_i + \sum_k W_{ik} \kappa(\zeta_k),
\end{equation}
where $b_i$ is the neuron's bias, $W_{ik}$ the synaptic weight and $\kappa(\zeta_k)$ the \gls{psp} induced by the last spike from neuron~$k$.

\newcommand{\vvzetak}{\vv{\zeta}_{\backslash k}}
The update rules of the Markov chain is determined by the transition operator $T(\vv{\zeta}'|\vv{\zeta})$ which describes the probability to obtain a certain state $\vv{\zeta}'$ given the current state~$\vv{\zeta}$.
The transition operator $\trop(\zeta_i'|\vv{\zeta})$ for a single neuron is given by:
\begin{subequations}\label{eq:methods:tr_op_single}
    \begin{alignat}{4}
        &\trop(\zeta'_i = 1 | \vv{\zeta}) &
        & =\left\{
            \begin{aligned}
                &r_i  \\
                &0 \qquad \\
            \end{aligned}\right. &\quad
        & \hspace{-0.5em}\begin{aligned}
            &\text{for } \zeta_i > \tauref \\
            &\text{else} \\
        \end{aligned} &
        &\quad \text{(spike)} \label{eq:methods:tr_op_single_1} \\
        &\trop(\zeta'_i = \zeta_i + 1 | \vv{\zeta}) &
        &= \left\{
            \begin{aligned}
                &1-r_i  \\
                &1 \qquad \\
            \end{aligned}\right. &\quad
        &\hspace{-0.5em}\begin{aligned}
            &\text{for } \zeta_i > \tauref \\
            &\text{else} \\
        \end{aligned} &
        &\hspace{.25em}\text{(no spike)} \label{eq:methods:tr_op_single_2}%
    \end{alignat}
\end{subequations}
All other transitions are forbidden.
As before, $\tauref$ represents the refractory period, but here as a natural number in terms of time steps.
We use a spiking probability $r_i = (1 + \exp(-(u_i - \log \tauref)))^{-1}$.
Because $u_i$ does not depend on $\zeta_i$ itself, all neurons can be updated in parallel.

\newcommand{\vvzetat}{\ensuremath{\vv{\zeta}^{(2)}}}
\newcommand{\zetaup}[1]{\ensuremath{\zeta_{#1}' = \zeta_{#1} + 1}}
\newcommand{\zetaone}[1]{\ensuremath{\zeta_{#1}' = 1}}
\newcommand{\givzeta}{\ensuremath{\zeta_1, \zeta_2}}
Starting from the transition rules \cref{eq:methods:tr_op_single} for one neuron, we can construct the transition rules for a two neuron system $\vvzetat = (\zeta_1, \zeta_2)$.
In this case, if neuron 1 is not refractory, it can spike with a probability of $r_1$ derived from $u_1 = b_1 + W_{12} \psp(\zeta_2)$.
Respectively, neuron 2 spikes with a probability determined by $u_2 = b_2 + W_{21} \psp(\zeta_1)$.
The resulting nine possible transitions are:
\begin{subequations}\label{eq:methods:trop2}
\begin{itemize}[leftmargin=*]
    \item neuron 1 and 2 are refractory ($\zeta_1 < \tauref$ and $\zeta_2 < \tauref$):
    \begin{align}\label{eq:methods:trop2_1}
        \trop(\zetaup{1}, \zetaup{2} | \givzeta) = 1
    \end{align}
    \item neuron 2 is refractory ($\zeta_2 < \tauref)$; 1 can spike ($\zeta_1 \ge \tauref$):
    \begin{align}
        &\trop(\zetaone{1}, \zetaup{2}, | \givzeta) &&= r_1\label{eq:methods:trop2_2}  \\ 
        &\trop(\zetaup{1}, \zetaup{2}, | \givzeta) &&= (1-r_1) \label{eq:methods:trop2_3}
    \end{align}
    \item neuron 1 is refractory ($\zeta_1 < \tauref)$; 2 can spike ($\zeta_2 \ge \tauref$):
    \begin{align}
        &\trop(\zetaup{1}, \zetaone{2},  | \givzeta) &&= r_2\label{eq:methods:trop2_4} \\
        &\trop(\zetaup{1}, \zetaup{2}, | \givzeta) &&= (1-r_2) \label{eq:methods:trop2_5}
    \end{align}
    \item neuron 1 and 2 can spike ($\zeta_1 \ge \tauref$ and $\zeta_2 \ge \tauref$):
    \begin{align}
        &\hspace{-1.5em}\trop(\zetaone{1}, \zetaone{2} | \givzeta) &&= r_1\,r_2\label{eq:methods:trop2_6} \\
        &\hspace{-1.5em}\trop(\zetaone{1}, \zetaup{2}, | \givzeta) &&= r_1\,(1-r_2)\label{eq:methods:trop2_7} \\
        &\hspace{-1.5em}\trop(\zetaup{1}, \zetaone{2}, | \givzeta) &&= (1-r_1)\,r_2\label{eq:methods:trop2_8} \\
        &\hspace{-1.5em}\trop(\zetaup{1}, \zetaup{2}, | \givzeta) \hspace{-1.2em}&&= (1-r_1)\,(1 - r_2) \hspace{-0.2em}\label{eq:methods:trop2_9}
    \end{align}
\end{itemize}
\end{subequations}
All other transitions are forbidden ($\trop(\zeta_1', \zeta_2' | \zeta_1, \zeta_2) = 0$).

With these transition probabilities (\cref{eq:methods:trop2}) we can compute \gls{stdd} of the two neuron system.
In the following derivation, we introduce a maximum value $Z \gg \max(\tauref, \tausyn)$ for $\zeta_i$ to limit the possible state space of our two neuron system to a finite number.

Let us define the probability state vector $\vv{\pi} \in \setReal^{Z^2}$, that contains the probabilities for all states,
\begin{equation}\label{eq:methods:pi}
    \vv{\pi} = \left( p(\vvzetat_{1,1} ), \dots, p(\vvzetat_{1, Z}), p(\vvzetat_{2, Z}), \dots, p(\vvzetat_{Z, Z}) \right)^\transp,
\end{equation}  
where we use the shorthand $\vvzetat_{i,j} = (\zeta_1 = i, \zeta_2 = j)$.
Together with the transition matrix $\mat{T} \in \setReal^{Z^2 \times Z^2}$
the evolution of the two neuron system from state \vv{\pi} to $\vv{\pi}'$ can be computed via
\begin{equation}
    \vv{\pi}' = \mat{T} \vv{\pi}.
\end{equation}
\mat{T} contains the transition probabilities defined in \cref{eq:methods:trop2} where the columns and rows are arranged in the same order as in \vv{\pi},

For calculating the \gls{stdd}, we are interested in the \emph{invariant} or \emph{steady state distribution} of the system, i.e. the distribution $\vv{\pi}^*$ which does not change when $\mat{T}$ is applied.
We obtain $\vv{\pi}^*$ by solving the eigenvalue problem
\begin{equation}
    \lambda \vv{\pi}^* = \mat{T} \vv{\pi}^*
\end{equation}
for the real eigenvalue $\lambda=1$.

In the following, we are calculating the probability that neuron 2 spikes $\DeltaT$ time steps after neuron 1 (i.e. the \emph{right} side of the \gls{stdd}).
To calculate negative $\DeltaT$ values (i.e. the left side), we can use the symmetry of the two neuron system and simply swap neuron 1 and 2 in the equations.

We start with a steady state vector $\vv{\pi}^*_\mathrm{right}$, where all entries for which $\zeta_1 > 1$ are set to zero, i.e. we consider only those states and their probabilities, for which neuron 1 has just spiked.

Obtaining a $\DeltaT$ means that both neurons undergo $\DeltaT - 1$ transitions where they increment their $\zeta$s (as described  by \cref{eq:methods:trop2_1,eq:methods:trop2_3,eq:methods:trop2_5,eq:methods:trop2_9}).
They are put together in the transition matrix $\mat{T}_\mathrm{nospike}$, which is basically a version of \mat{T} in which all rows in which $\zeta_1' = 1$ or $\zeta_2' = 1$.

After the $\DeltaT - 1$ steps, neuron 2 spikes for which we construct an additional matrix $\mat{T}_\mathrm{spike} \in \setReal^{Z \times Z^2}$ that contains the transition probabilities \cref{eq:methods:trop2_4,eq:methods:trop2_8}.
Hence, the right side of the \gls{stdd} $\vv{p}(\DeltaT>0)$ is given by
\begin{equation}
    \vv{p}(\DeltaT>0) = \sum_{i = 1}^{Z - 1} \mat{T}_\mathrm{spike} \mat{T}_\mathrm{nospike}^{i - 1} \vv{\pi}^*_\mathrm{right}.
\end{equation}
The whole distribution is obtained by concatenating the right and left distribution.

\subsubsection{Proof for the existence and stability of fixed points in SAL}
\label{sec:methods:proof}

To show the existence and stability of fixed points in \gls{sal}, we introduce the following notation:
We use a $\hat{\cdot}$ to indicate the time-mirrored version of a variable, i.e., $\hat{g}(t) \coloneqq g(-t)$.
Furthermore, we use a $\cdot^+$ to indicate the positive part of a function, i.e. positive domain $t\ge0$, and $\cdot^-$ for the negative part.

For the \gls{stdp} kernel %
we assume $\stdpwindowCausal(t) > 0$, $\stdpwindowAntiCausal(t) < 0$ and $\stdpwindowCausal(t) = -\hat{\stdpwindow}^-(t)$ as well as positive derivatives ${\stdpwindowCausal}^{\prime}(t) = {\stdpwindowAntiCausal}^{\prime}(t) > 0$. %
The exact shape does not matter in our consideration.

We denote the \gls{stdd} for a given set of symmetric weights $W_{12} = W_{21}$ with $p_0(t)$.
Additionally, we introduce a distribution $p_\varepsilon(t)$ of a set of weights $W_{21} = W_{12} + \epsilon$ with a small $\varepsilon \to 0$.
From this, we define for $t \ge 0$ the deviation
\begin{equation}\label{eq:methods:def_delta}
    \delta \coloneqq p^+_\varepsilon - p^+_0 - \left( \hat{p}_\varepsilon^- - \hat{p}^-_0 \right).
\end{equation}
Importantly, the \gls{stdd} is defined such that it contains only nearest-neighbor spike pairs.
Hence, every causal spike pair is followed by an anti-causal pair, and it follows directly that $\int\limits_0^\infty p^+(t) \dd t = \int\limits_0^\infty \hat{p}^-(t) \dd t$%
Using the definition of the deviation \cref{eq:methods:def_delta}, we thus note that
\begin{equation}\label{eq:methods:int_delta}
    \int\limits_0^\infty \delta(t) \dd t = 0.
\end{equation}

\begin{figure}[t]
    \centering
    \begin{tikzpicture}
        \node[anchor=north west] at (0.0, 0.0) {\includegraphics[width=4cm]{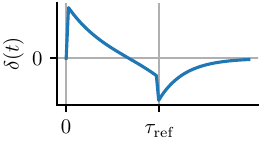}};
        \node[anchor=north west] at (4.0, 0.0) {\includegraphics[width=4cm]{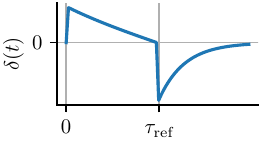}};
        \node[anchor=north west] at (0.0, -0.0) {\figanno{a}};
        \node[anchor=north west] at (0.0, -2.3) {\includegraphics[width=4cm]{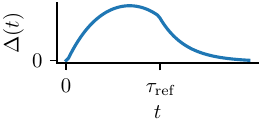}};
        \node[anchor=north west] at (4.0, -2.3) {\includegraphics[width=4cm]{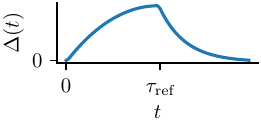}};
        \node[anchor=north west] at (0.0, -2.0) {\figanno{b}};
    \end{tikzpicture}
    \caption{
        \captionanno{a} \gls{stdd} deviation $\delta(t)$ for $W_{ij} = W_{ji} + \varepsilon$ (with $\varepsilon = 0.005$) and rectangular \glspl{psp}, for positive (left) and negative weights (right). \captionanno{b}  Integrated deviation $\Delta(t)$ for positive (left) and negative (right) weights.
    }
    \label{fig:methods:proof_delta}
\end{figure}

To prove the existence of fixed points of \gls{sal}, we turn to the expectation value for a weight update $\langle \dot{W}_{ij} \rangle$.
The expectation value for a symmetric \gls{stdp} kernel is given by
\begin{align}\label{eq:methods:exp_val}
    \big\langle \dot{W}_{ij} \big\rangle &=  \int\limits_{-\infty}^0 f^-(t) p^-(t) \dd t + \int\limits_0^\infty f^+(t) p^+(t) \dd t \\
    &= \int\limits_0^\infty f^+(t) \left[ p^+(t) - \hat{p}^-(t) \right] \dd t.
\end{align}
For brevity, we omit the synapse indices here for the \gls{stdd} $p_{ij}$.
    Assuming a symmetric \gls{stdd} for symmetric weights $W_{ij} = W_{ji}$, we have $p_0^+ = \hat p_0^-$, so the average weight update $\Delta \dot W_{ij} = 0$. 
    Thus, $W_{ij} = W_{ji}$ is a unique fixed point.

We now show that the fixed points are stable under small perturbations of one weight, which is represented by $p_\varepsilon$.
That means that $\langle \dot{W}_{ji} \rangle < 0$ if $\varepsilon > 0$, i.e., $W_{ji} > W_{ij}$ and vice versa that $\langle \dot{W}_{ji} \rangle > 0$ if $\varepsilon < 0$.
Since the \glspl{stdd} are identical (just time-mirrored) for the two synapses, the same derivation holds for $W_{ij}$.
We start with $\varepsilon > 0$.
The expectation value analog to \cref{eq:methods:exp_val} is
\begin{align}\label{eq:methods:exp_val_eps}
    \big\langle \dot{W}_{ji} \big\rangle &= \int\limits_0^\infty f^+(t) \left[ p^+_\varepsilon(t) - \hat{p}^-_\varepsilon(t) \right] \dd t \\
    &= \int\limits_0^\infty f^+(t) \left[ p^+_0(t) + \delta(t) - \hat{p}^-_0(t) \right] \dd t \\
    &= \int\limits^\infty_0 f^+(t) \delta(t) \dd t.
\end{align}
By introducing the variable $\Delta(t) \coloneqq \int\limits_0^t \delta(t^\prime) \dd t^\prime$ and integrating by parts, we can reformulate \cref{eq:methods:exp_val_eps} as
\begin{equation}
    \big\langle \dot{W}_{ji} \big\rangle = f^+(t) \Delta(t) \Bigg\rvert^\infty_0 - \int\limits_0^\infty f^{+\prime}(t) \Delta(t) \dd t.
\end{equation}
Because of $\Delta(0) = 0$ and $\lim_{t \to \infty} \Delta(t) = 0$  (see \cref{fig:methods:proof_delta}), the boundary terms vanish.
Importantly, because of $\Delta(t) > 0$ for all $t > 0$ and because of our assumption ${\stdpwindowCausal}^\prime(t) > 0$, the integral evaluates to a finite positive number.
Therefore, for a positive $\varepsilon$, we have $\langle \dot{W}_{ji} \rangle < 0$.
For a negative $\varepsilon$, $\delta$ and consequently $\Delta$ change the sign and therefore the integral term evaluates to a negative number, and $\langle \dot{W}_{ji} \rangle > 0$.
Hence, in the vicinity around the fixed point $W_{ji} = W_{ij}$, the weight updates move $W_{ji}$ towards $W_{ij}$, which shows that $W_{ij} = W_{ji}$ is a stable solution.

In \cref{sec:sm:stdd_ana_vs_sim}, we demonstrate the agreement between the analytical method presented here and \glspl{stdd} derived from numerical simulations.

\subsection{Boltzmann Machines/Sampling with spikes}
\label{sec:methods:snn}

To demonstrate the effectiveness of \gls{sal} in networks with strong recurrence, we turn to \glspl{ssn}, which we train to represent arbitrary Boltzmann distributions. 
The framework of \glspl{ssn} was introduced in~\cite{buesing2011neural}.
It builds on the theory of \glspl{bm}, which originate from the field of statistical physics and are used to represent probability distributions over binary variables $z_i$ of a system of $N$ coupled spins or -- in the language of neural networks -- a recurrent network of stochastic binary neurons.

For this system, we define an energy function
\begin{equation}\label{eq:methods:bm_energy}
    E(\vv{z}) = - \frac{1}{2} \vv{z}^\transp\mat{W}\vv{z} - \vv{b}^\transp\vv{z},
\end{equation}
that assigns a scalar energy value to every possible configuration $\vv{z} \in [0, 1]^N$ of the system.
$\mat{W} \in \setReal^{N \times N}$ is a symmetric weight matrix with vanishing diagonal that contains the interactions strengths between the neurons and $\vv{b} \in \setReal^N$ the vector of neuronal biases.

The probability of a state \vv{z} is determined by the Boltzmann distribution,
\begin{equation}\label{eq:methods:bm_distribution}
    P(\vv{z}) = \frac{1}{Z} \exp \left( -E(\vv{z}) \right),
\end{equation}
where $Z = \sum_{\vv z'} \exp \left( -E(\vv z) \right)$ is the partition function.

To make the link between the theory of \glspl{bm} and spiking neural networks, we map the refractoriness of a neuron to the binary variable $z_i$, where $z_i = 1$ is assigned to a neuron which is refractory and $z_i = 0$ to a neuron which is not.
The neuron model employed here is the same as defined in \cref{sec:methods:neuronmodel} by \cref{eq:methods:mempot_gls,eq:methods:rect_psp,eq:methods:probspk,eq:methods:act_func}.
Driven by its intrinsic randomness, the network randomly evolves over time and explores different network states along some random trajectory through the state space, which we call sampling.
The relative frequency by which the states $z$ are visited represents the underlying joint probability distribution $p(\vv{z})$.
In \cite{buesing2011neural}, the authors have shown the equivalence between \gls{mcmc} from Boltzmann distributions and the sampling with networks of spiking \glspl{glm}.

By choosing a suitable number of neurons and partitioning the network into a population of \emph{visible} and \emph{hidden} neurons, the marginal distribution over the visible population can approximate any joint probability distribution.

The training of the network consists of minimizing the difference between the network distribution $p(\vv{z})$ and the target distribution $p^*(\vv{z})$.
To do so, we minimize the \gls{dkl}, a standard measure of similarity between two probability distributions.
Minimization with respect to the network's weights and biases yields optimized parameters
\begin{equation}\label{eq:methods:argmax_dkl}
    \hat{\vv \theta} = \argmin_{\vv \theta} \gls{dkl}(p\|p^*), \quad \vv \theta \in \{ \vv b, \mat W \},
\end{equation}
where $\gls{dkl}(p\|p^*) \coloneqq \sum_{\vv{z}} p(\vv z ) \log \frac{p(\vv z)}{p^*(\vv z)}$.
The weights and biases are updated iteratively by performing gradient descent on \gls{dkl}.
Training is split into two reoccurring phases:
In the \emph{wake phase}, the network is constrained by the target distribution (it \enquote{sees} the target pattern), while during the \emph{sleep phase}, it is allowed to sample freely from its internal distribution.
This results in the wake-sleep algorithm used for training \glspl{bm} \cite{ackley1985learning}%
\begin{align}
    \Delta W_{ij} &= \eta_W \left[ \langle z_i z_j \rangle_\mathrm{wake} - \langle z_i z_j \rangle_\mathrm{sleep}\right] \label{eq:methods:wake_sleep_W} \\
    \Delta b_i &= \eta_b \left[ \langle z_i \rangle_\mathrm{wake} - \langle  z_i \rangle_\mathrm{sleep} \label{eq:methods:wake_sleep_b} \right],
\end{align}
where $\langle \cdot \rangle_\mathrm{wake} = \langle \cdot \rangle_{p^*(\vv z)}$ denotes the expectation value with respect to the target distribution $p^*(\vv z)$ and $\langle \cdot \rangle_\mathrm{sleep} = \langle \cdot \rangle_{p(\vv z | \vv b, \mat W)}$ with respect to the distribution of the free model.
We refer to \cref{eq:methods:wake_sleep_W,eq:methods:wake_sleep_b} as \emph{state-based} wake-sleep.

In an \gls{ssn}, the estimation of the correlations $\langle z_i z_j \rangle_x$ can be realized directly on the spike-trains by a standard \gls{stdp}-rule as introduced in \cref{eq:methods:stdp}, which we refer to as \emph{spike-based} wake-sleep. 
The \gls{stdp}-window has the shape of a left-right symmetric triangle with $f(\DeltaT) = \max\left(-\frac{\DeltaT}{\tauref} - 1, 0\right)$ and $\stdpBoth =  1$ for the wake and $\stdpBoth =-1$ for the sleep phase.
In the limit of small learning rates, the accumulated weight updates using spike-based \gls{stdp} after the two phases is equivalent to the weight update produced by state-based wake-sleep.
This allows us to minimize $\gls{dkl}$ in an efficient online manner, relying only on the knowledge of last spike times instead of accumulated expectation values.

\subsubsection{Simulation details}

We train a fully connected \gls{bm} of size $N=7$ to sample from random target distributions of the same dimension.
To compute the target distributions $p^*$, we generate Boltzmann distributions using \cref{eq:methods:bm_distribution} with parameters $\vv{W}^*$ and $\vv{b}^*$ sampled from a uniform distribution.
This ensures that the \gls{bm} is able to solve the task of representing $p^*$ with high precision.

Prior to training, the network is initialized with random biases $\vv{b}_\mathrm{init}$ and weights $\mat{W}_\mathrm{init}$ drawn from a normal distribution.
The upper triangle of $\vv{W}_\mathrm{init}$ is copied to the lower triangle to ensure symmetry. 

Depending on the experiment type, parameter noise is added to the network upon initialization:
In the \emph{synaptic noise scenario}, Gaussian noise with $\mu=0$ and a standard deviation $\noiseinit$ is added to each weight to model weight asymmetry of different strength prior to training.
In the \emph{plasticity noise scenario}, all runs are conducted with Gaussian noise with $\noiseinit = 0.2$ added to $\mat W_\mathrm{init}$.
Additionally, we also add noise to the \gls{stdp} factors:
For each synapse $W_{ij}$, a random noise value $\xi_{ij}$ is drawn from a normal distribution with $\mu=0$ and $\noisestdp$.
For the wake phase, this is added to both causal and anticausal prefactors, $\stdpBoth + \xi_{ij}$, and subtracted for the wake phase, $\stdpBoth - \xi_{ij}$.
This models synaptic heterogeneity and ensures that reciprocal synapses produce non-equal weights updates, although they receive the same spike trains and should produce as the same updates as expected.

Each scenario is simulated twice, one time without \gls{sal} and a second time with an additional \gls{sal} phase.
To accelerate learning, sampling during the wake phase is replaced by calculated target coactivation $\langle z_i z_j \rangle_\mathrm{wake}$ and rates $\langle z_i \rangle_\mathrm{wake}$.
Weight updates are applied in batches after each phase (wake-sleep or \gls{sal}).
Learning rates are optimized through visual inspection, and training progress is monitored using the \gls{dkl} between $p$ and $p^*$.
For each scenario, a reference run without the specific noise type is conducted (blue/purple lines and markers in \cref{fig:results:snn_init_noise,fig:results:ssn_stdp_noise}), and training is halted when the \gls{dkl} ceases to decrease.
For each noise level in \cref{fig:results:snn_init_noise,fig:results:ssn_stdp_noise}, we train the same 5 distributions with 4 seeds each, resulting in different 20 runs, and compare the \gls{dkl} and variance of weight differences $\var \left( W_{ij} - W_{ji} \right)$ between runs with and without \gls{sal}.
The relevant simulation parameters are given in \cref{tab:methods:ssn_simparams}.

\begin{table*}
    \caption{
        Simulation parameters for the \gls{ssn} experiments.
    }
    \label{tab:methods:ssn_simparams}
    \centering
    \begin{tabular}{lllllll}
\toprule
                                    & \multicolumn{3}{c}{synaptic noise}                  & \multicolumn{3}{c}{plasticity noise}                \\
                                    & w/o SAL         & w/ SAL          & KP              & w/o SAL         & w/ SAL          & KP              \\ \midrule
$\mat{W}^*$                         & \multicolumn{6}{c}{$\mathcal{U}[-1, 1]$}                                                                  \\
$\vv{b}^*$                          & \multicolumn{6}{c}{$\mathcal{U}[-1, 1]$}                                                                  \\
$\mat W_\mathrm{init}$              & \multicolumn{6}{c}{$\mathcal{N}(0, 0.2)$}                                                                 \\
$\vv b_\mathrm{init}$               & \multicolumn{6}{c}{$\mathcal{N}(0, 0.2)$}                                                                 \\
$\noiseinit$                        & \small variable & \small variable & \small variable & 0.2             & 0.2             & 0.2             \\
$\noisestdp$                        & 0               & 0               & 0               & \small variable & \small variable & \small variable \\
$\eta_{\mat{W}}$                    & 0.01            & 0.01            & 0.01            & 0.001           & 0.001           & 0.001           \\
$\eta_{\vv{b}} $                    & 0.01            & 0.01            & 0.01            & 0.001           & 0.001           & 0.001           \\
$\eta_\mathrm{SAL} $                & 0               & 0.01            & 0               & 0               & 0.002           & 0               \\
$\lambda$                           & 0               & 0               & \num{4e-5}      & 0               & 0               & \num{4e-5}      \\
simulation time step                & \multicolumn{6}{c}{\SI{0.2}{\milli\second}}                                                               \\
$\tauref$                           & \multicolumn{6}{c}{\SI{10}{\milli\second}}                                                                  \\
\small duration sleep phase & \multicolumn{6}{c}{\SI{10}{\second}}                                 \\ \bottomrule
    \end{tabular}
\end{table*}

\subsection{Spiking cortical microcircuits}
\label{sec:methods:mc}

To demonstrate \gls{sal} in the context of biologically plausible \gls{bp} in spiking networks, we turn to the model of cortical microcircuits presented in~\cite{sacramento2018dendritic} and extend it to communication with actual spikes instead of rates.

\subsubsection{Mathematical description of the model}
\label{sec:methods:mc_math}

The general scheme of the circuit is depicted in \cref{fig:results:mc}.
For a proof of the equivalence between the (rate-based) microcircuit model and \gls{bp}, we refer the reader to \cite{sacramento2018dendritic} and \cite{max2024learning}.

In the absence of a teaching signal, e.g., from higher cortical areas, pyramidal cells function as representation units, responsible for feedforward activation.
We denote variables associated with these cells with a superscript $\cdot^\mathrm{P}$.
They receive bottom-up sensory inputs via their basal dendrites and top-down error/learning signals through the apical dendrite, integrating these inputs at the soma.
Such preferential targeting of dendritic compartments by top-down/bottom-up projections is consistent with observations of pyramidal cells in cortex~\cite{petreanu2009subcellular,larkum2013cellular,makino2015learning,gillon2024responses,jordan2020opposing}.
In the model, the complex dynamics of pyramidal cells are modelled with a simplified three-compartment model, which includes distinct basal, apical, and somatic voltages. 

Interneurons (indicated by a superscript $\cdot^\mathrm{I}$) are located in the hidden layers of the network and aim to replicate the activation of pyramidal cells in the layer above.
They are divided into two compartments, representing the dendritic tree and the soma.
Across the layers, the populations of pyramidal neurons and interneurons are organized such that the number of interneurons in the hidden layers matches the number of pyramidal cells in the layer above.
Pyramidal cells project laterally to these interneurons and receive signal back from them to their apical tree.

In accordance with the previous models of this work (\cref{sec:methods:neuronmodel,sec:results:ssn}), neurons are modelled as \glspl{glm}, i.e., we use the spiking mechanism presented in \cref{eq:methods:probspk} together with the activation function \cref{eq:methods:act_func}.
The membrane potential dynamics of pyramidal cells integrate the compartments,
\begin{equation}\label{eq:methods:pyr_mempot}
    \vuP_{\ell}(t) = \vv{b}_{\ell} + \lamBas \vvBas_{\ell}(t) + \lamApi \vvApi_{\ell}(t),
\end{equation}
where $\vv{b}_{\ell}$ represents the neuronal biases and $\vvBas_{\ell}$ and $\vvApi_{\ell}$ the basal and apical voltages respectively together with their coupling strengths \lamBas and \lamApi.

The basal compartment $\vvBas_{\ell}(t) = \mWPP_{\ell, \ell-1} \vpsp^\mathrm{p}_{\ell-1}(t)$ receives input spikes from the pyramidal neurons in the layer below. 
Here, $\mWPP_{\ell, \ell-1}$ denotes the bottom-up weights from layer $\ell-1$ to $\ell$ and $\hat{\psp}^x_{\ell-1, i}(t) = \int _{0}^{\infty} \psp(s)\spktrain^x_{\ell-1, i}(t - s) \, \dd s$ the filtered spike train from neuron $i$ in layer $\ell-1$.

The apical compartment $\vvApi_{\ell}(t) = \mBPP_{\ell, \ell+1} \vpsp^\mathrm{p}_{\ell+1}(t) + \mWPI_{\ell, \ell} \vpsp^\mathrm{I}_{\ell}(t)$ integrates the top-down input from the upper layer (through $\mBPP_{\ell, \ell+1}$) and compares it to the activity of the interneurons received through the lateral weights $\mWPI_{\ell, \ell}$.

In this model, interneurons consist of two compartments, representing the soma and a dendritic tree.
Their somatic voltage is given by
\begin{equation}\label{eq:methods:in_mempot}
    \vuI_\ell(t) = \vv{b}_\ell + \lamDen \vvDen_\ell(t) = \vv{b}_\ell + \lamDen \mWIP \vpsp^\mathrm{P}_\ell(t),
\end{equation}
receiving input from the population of pyramidal neurons in the same layer through afferent lateral weights \mWIP.

During training, the top layer neurons are described by
\begin{equation}\label{eq:methods:pyr_mempot_tl}
    \vuP_L(t) = \vv{b}_L + \lamBas \vvBas_L(t) + \lamNudge \vvApi_L(t),
\end{equation}
where the apical compartment induces a nudging by the current error signal, i.e.~the difference between the target voltage $\vuTgt$ and bottom-up input plus bias, $\vvApi_L = \vuTgt - (\vv{b}_L + \lamBas \bvvBas_L)$.
Note that we have introduced a time-smoothed version $\bvvBas_L$ of the basal input, which we obtain by taking a moving average over present and past voltages~$\vvBas_L(t)$.
The averaging of $\vvBas$ is needed because the target is provided by a vector with smooth, continuous values and hence the error signal encoded in $\vvApi$ is required to be smooth in time as well.
$\lamNudge$ controls the nudging strength of the target, which is set to zero in absence of a teaching signal.

The microcircuit model has to be operated in the so-called \emph{self-predicting state}, in which $\vuI_\ell$ matches $\vuP_{\ell}$ in the absence of a top-down teaching signal. 
In this state, the apical voltage $\vvApi_\ell$ is zero when the network receives no training signal (or the network has learned perfectly) because the top-down input from the layer $\ell +1$ and the lateral input from the interneuron in layer $\ell$ cancel.
The self-predicting state is realized if $\mWIP_{\ell,\ell} = \frac{\lamBas}{\lamDen} \mWPP_{\ell+1, \ell}$, which can be dynamically learned as done in the original model.
Here, we set it for computational efficiency.

Following this logic, $\mWPI_{\ell, \ell}$ and $\mBPP_{\ell, \ell+1}$ are matched such that $\vvApi_{\ell} = 0$ in absence of a top-down teaching signal.
This, too, can be achieved dynamically by a local learning rule~\cite{sacramento2018dendritic}; here, we also set $\mWPI_{\ell, \ell} = - \mBPP_{\ell, \ell+1}$.

Following the derivation presented in \cite{sacramento2018dendritic}, one can show that $\vvApi_{\ell}$ effectively encodes a local error signal: 
If $\vuP_{\ell+1}$ is nudged towards a target, a residual voltage is induced in $\vvApi_\ell$.
It can be shown that this model together with a local plasticity rule inspired by \cite{urbanczik2014learning} can approximate the \gls{bp} algorithm \cite{sacramento2018dendritic, max2024learning}.
Our spiking adaptation of the bottom-up learning rule connecting neuron $i$ to neuron $j$ is
\begin{align}\label{eq:methods:mc_lr}
    \dot{W}^\mathrm{PP}_{\ell, \ell-1, \, ji}(t)\, &= \\
    \eta \, \Big[ \varphi(\buP_{\ell, j}&(t)) - \varphi(b_{\ell, j} + \lamBas \bvBas_{\ell, j}(t)) \Big] \, \spktrain^\mathrm{P}_{\ell-1, i}(t) \,,\nonumber
\end{align}
where $\eta$ is the learning rate and $\varphi(u) = \tauref^{-1} \exp (u)$ the activation function of the \gls{glm}.
In addition to $\bvBas$, the learning rule also requires the smoothed somatic membrane potential $\buP$.

Similar to \gls{pal} for the rate-based case~\cite{max2024learning}, we endow the (now spiking) microcircuit model with \gls{sal}, replacing \gls{fa} with dynamical alignment of backward connections $\mBPP$ with their feedforward partner weights $\mWPP$.

\subsubsection{Simulation details}
\label{sec:methods:mc_exp}

Using this setup, we demonstrate that \gls{sal} outperforms \gls{fa}, and retains the efficient credit assignment of \gls{bp}.
We point out that the simulations are carried out with fully recurrent dynamics and fully spike-based communication as described by \crefrange{eq:methods:pyr_mempot}{eq:methods:mc_lr}, setting our work apart from similar approaches, where spikes are replaced by rates and neuronal dynamics by steady-state approximations, or the challenges of recurrence are lifted by computing the forward and backward passes separately.

We use a teacher microcircuit network to produce a nonlinear input-output mapping that the student network has to learn.
The teacher consists of one hidden microcircuit and a pyramidal output neuron.
Teacher and student have the same size and parametrization, but only the student is nudged by the target.
Crucially, to achieve the optimal output, the student network has to learn the exact same weights as the teacher, which is only possible if a meaningful error signal arrives at the hidden neuron.

Training is divided into epochs of 20~shuffled training inputs, an optional \gls{sal} phase of five inputs, and a validation phase without nudging of six validation inputs.
Each input consists of a voltage that is converted into spike trains using \cref{eq:methods:act_func,eq:methods:probspk}. The input for the training phases consists of nine equally spaced voltages between $-3$ and $3$; the input for the validation consists of six voltages between $-2.7$ and $2.7$.
In all cases, the targets are the time-averaged somatic voltages~$\uP_L$ recorded from the output neuron of the teacher network.

When using \gls{bp}, the value of $\WPP_{2,1}$ is copied to $\BPP_{1,2}$ after every time step.
When using \gls{sal}, $\lamApi$ is set to 1 during the \gls{sal} phases.
Otherwise, \gls{sal} plasticity is not activated in $\BPP_{1,2}$.

All relevant simulation parameters are given in \cref{tab:methods:mc_simparams}.

\begin{figure}[htb]
    \centering
    \includegraphics[width=\linewidth]{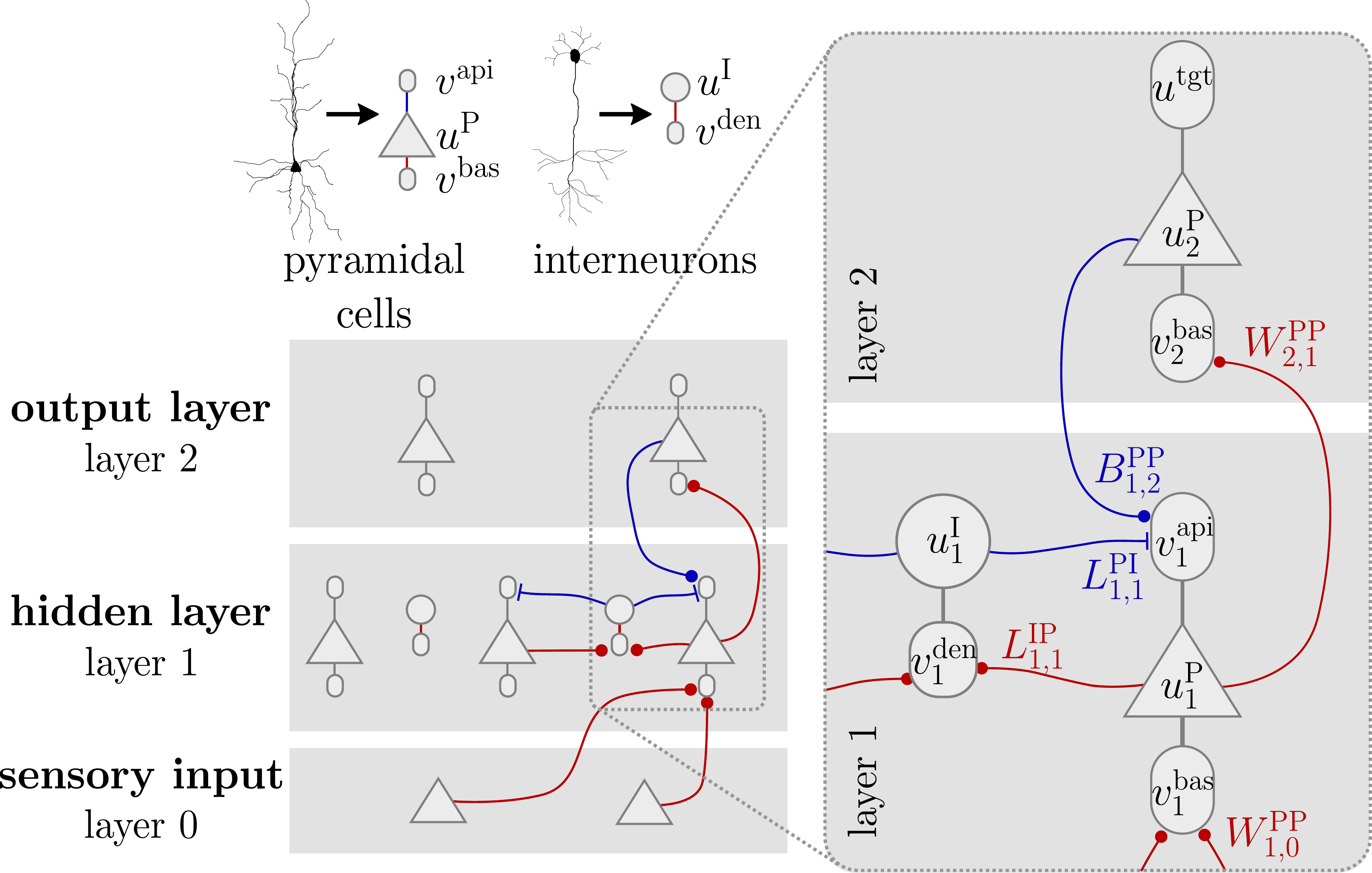}
    \caption{Schematic of the microcircuit model: bio-plausible transportation of error signals and local error representation in apical dendrites. Adapted from~\cite{max2024learning}}
    \label{fig:methods:mc}
\end{figure}

\begin{table}[htb]
    \centering
    \caption{
        Simulation parameters for the cortical microcircuits.
    }
    \label{tab:methods:mc_simparams}
    \begin{tabularx}{\linewidth}{Xccc}
    \toprule
                & BP        & FA       & SAL       \\ \midrule
    \lamApi     &   \multicolumn{3}{c}{0.02}       \\
    \lamNudge   &   \multicolumn{3}{c}{0.6} \\
    \lamBas     &   \multicolumn{3}{c}{1.0}       \\
    \lamDen     &   \multicolumn{3}{c}{1.0}       \\
    simulation time step  &  \multicolumn{3}{c}{\SI{1}{\milli\second}} \\
    $t_\text{moving average}$ & \multicolumn{3}{c}{\SI{2}{\second}} \\
    $\tauref$   &  \multicolumn{3}{c}{\SI{10}{\milli\second}} \\
    $\eta$ for $\WPP_{1,0} $ &  0.2 & 0.2 & 0.2 \\
    $\eta$ for $\WPP_{2,1} $ &  0.003 & 0.003 & 0.003 \\
    $\eta$ for $\BPP_{1,2} $ &  n.a. & 0.0 & 0.001 \\
    $b_1$       & \multicolumn{3}{c}{$-1.0$} \\
    $b_2$       & \multicolumn{3}{c}{$-1.0$} \\
    teacher $\WPP_{1, 0}$ & \multicolumn{3}{c}{2.0} \\
    teacher $\WPP_{2, 1}$ & \multicolumn{3}{c}{2.0} \\
    training values \newline per iteration & \multicolumn{3}{c}{20} \\
    validation values \newline per iteration & \multicolumn{3}{c}{6} \\
    values for SAL \newline per iteration & 0  & 0  & 5 \\
    presentation time & \multicolumn{3}{c}{\SI{20}{\second}} \\
    student init $\WPP_{1,0}$  & \multicolumn{3}{c}{$\mathcal{U}[-3, 3]$} \\
    student init $\WPP_{2,1}$  & \multicolumn{3}{c}{$\mathcal{U}[-3, 3]$} \\
    student init $\BPP_{1,2}$  & $=\WPP_{2,1}$ & $\mathcal{U}[-3, 3]$ & $\mathcal{U}[-3, 3]$ \\
    \bottomrule
    \end{tabularx}
\end{table}

\subsection{Deep symmetrization network (SymmNet)}
\label{sec:methods:symmnet}

\subsubsection{Preliminary study with two layers}

The design consists of two bidirectionally densely connected layers.
The goal is to align the backward weights $\mat B$ with the fixed forward weights $\mat W$.
The initial forward and backward weights $\mat W$ and $\mat B$ are independently drawn from a uniform distribution according to the Kaiming He initialization scheme~\cite{he2016deep}.
The neuron and synapse model for the run with \gls{sal} is identical with the experiments before, i.e., the \gls{glm} with rectangular \gls{psp} shapes.
The implementations of \gls{rdd}~\cite{guerguiev2019spikebased} and \gls{stdwi}~\cite{ahmad2020overcoming} used here are identical to the models described in the respective original publications.
We used the original code base published alongside the original publications with minimal changes.
In particular, because the publication for \gls{stdwi}~\cite{ahmad2020overcoming} did not specify all relevant settings and parameter, we used the parameters found in the code base\footnote{
    \url{https://github.com/nasiryahm/STDWI/tree/master}, commit SHA \texttt{1d6788a0e38da49cdfa59c4ed4bcd5f00ec0462c}
}.

The learning rates for \gls{sal} and \gls{stdwi} are tuned by hand to achieve the best alignment angles within a convergence time of \SI{1000}{\second}.

\begin{table}[htb]
\caption{Simulation parameters for the preliminary two layer study}
\label{tab:methods:stdwi}
\begin{threeparttable}
\begin{tabularx}{\linewidth}{Xc}
\toprule
                          & \scshape Common Settings      \\ \midrule
input size                & 50                            \\
output size               & 50                            \\
$\mat W_\mathrm{init}$    & uniform Kaiming He \cite{he2016deep} \\
$\mat B_\mathrm{init}$    & uniform Kaiming He \\ \midrule
                          & \scshape SAL                  \\ \midrule
simulation time step      & \SI{1}{\milli\second}         \\
learning rate             & 0.05                          \\
$\tauref$                    & \SI{10}{\milli\second}        \\ \midrule
                          & {\scshape RDD}\tnote{a}                  \\ \midrule
stimulation ratio         & 0.2                           \\
RDD time                  & \SI{1}{\second}                \\ \midrule
                          & \scshape STDWI                \\ \midrule
simulation time step      & \SI{0.25}{\milli\second}      \\
stimulation ratio         & 0.2                           \\
stimulation correlation   & 0.0                           \\
stimulation rate          & \SI{200}{\hertz}              \\
stimulation duration      & \SI{100}{\milli\second}       \\
stimulation weight scale  & 10                            \\
weight scale              & 75                            \\
STDP $\tau_\mathrm{fast}$ & \SI{10}{\milli\second}        \\
STDP $\tau_\mathrm{slow}$ & \SI{200}{\milli\second}       \\
learning rate             & 0.02                          \\
decay weighting           & 0.1                           \\ \bottomrule 
\end{tabularx}

\begin{tablenotes}[flushleft]   %
  \item[a] {\small All \gls{rdd}-specific model parameters are the same as in the original publication~\cite{guerguiev2019spikebased}.}
\end{tablenotes}
\end{threeparttable}

\end{table}

\subsubsection{SymmNet experiments}

The network architecture of the \gls{symmnet} is identical to the design in \cite{guerguiev2019spikebased}.
It consists of two different but linked networks: 
a classical (non-spiking) \gls{ann} for the training of the forward weights $\mat{W}_\ell$ and for inference, which we refer to as \gls{convnet}, and a complementary \gls{snn} for symmetrization, i.e., the training of the backward weights $\mat{B}_\ell$.

The \gls{convnet} consists of two convolutional blocks, each composed of a 2D convolutional layer with ReLU activation function, 2D max-pooling and 2D batch normalization. 
On top of that, there are three \gls{fc} feedforward layers with ReLU activations and 1D max-pooling (see \cref{fig:results:symmnet}\crefanno{a} and \cref{tab:methods:symmnet}).

The spiking network matches the architecture of the \gls{convnet}'s \gls{fc} block.
For the \gls{salnet}, the three layers consist of \gls{glm} neurons as described in \cref{sec:methods:neuronmodel}.
The bottom layer of \gls{salnet} does not receive any external input but is driven only by the internal noise.
For the \gls{rddnet}, the neurons implement the \gls{rdd} mechanism as described in~\cite{guerguiev2019spikebased}.
For both \gls{sal} and \gls{rdd}, only weights $\mat B_\ell$ are plastic, learning to approximate $\mat W_\ell^\transp$.

Weights $\mat W$ and biases $\vv b$ of the \gls{convnet} are trained with stochastic gradient descent by minimizing the cross-entropy between the actual and predicted class of the data using standard autograd methods.
To summarize, we employ the following training schemes:
\begin{itemize}
    \item normal \textbf{\gls{bp}}, where $\mat B_\ell \mapsfrom \mat W_\ell^\transp$
    is enforced after every weight update,
    \item \textbf{\gls{fa}}, where $\mat B_\ell$ is initialized randomly and kept fixed,
    \item \textbf{weight decay} (Kolen~\& Pollack \cite{kolen1994backpropagation}, Akrout et al. \cite{akrout2019deep}, eq.~16 and 17) and
    \item \textbf{\gls{sal}} and \textbf{\gls{rdd}}, where the weights and biases are copied from the \gls{convnet} into the \gls{salnet} (\gls{rddnet}) before every epoch, the backward weights are learned for a specific time and copied back into the \gls{convnet}. 
    During each epoch of $\mat W_\ell$-training with stochastic gradient descent, $\mat B_\ell$ is kept fixed as in \gls{fa}.
\end{itemize}
Importantly, the convolutional layers are trained with \gls{fa} in all cases except for \gls{bp} due to biologically implausible requirement for weight sharing in convolutional layers.
This explains the remaining performance gap with respect to \gls{bp}.

We limit the number of epochs to 200 for all training algorithms, after which \gls{bp} has converged, and report the final validation accuracy.
While \gls{kp} symmetrization benefits from minuscule learning and decay rates, we chose this stopping point for practical feasibility and to ensure a fair comparison of the different symmetrization methods.

\begin{table*}[htbp]
\caption{Simulation details for the \gls{symmnet} experiment}
\label{tab:methods:symmnet}

\begin{threeparttable}
\begin{tabularx}{\linewidth}{lXccX}
\toprule
Symbol                                        & Parameter name                & CIFAR-10                           & SVHN                           & Fashion-MNIST                           \\ \midrule
\multicolumn{5}{c}{\scshape\gls{convnet} architecture}\\ \midrule
                                              & input size                    & \multicolumn{2}{c}{$32 \times 32 \times 3$} & $28 \times 28 \times 1$ \\
                                              & \multirow{2}{*}{conv layer 1} & \multicolumn{3}{c}{Conv2D $(5\times5)\times 64$, ReLU}\\
                                              &                               & \multicolumn{3}{c}{MaxPool2d (kernel size = 2, stride = 2)}\\
                                              & \multirow{2}{*}{conv layer 2} & \multicolumn{3}{c}{Conv2D $(5\times5)\times 64$, ReLU}\\
                                              &                               & \multicolumn{3}{c}{MaxPool2d (kernel size = 2, stride = 2)}\\
                                              & FC layer 1                    & \multicolumn{3}{c}{size\,384, ReLU}\\
                                              & FC layer 2                    & \multicolumn{3}{c}{size\,192, ReLU}\\ 
                                              & FC layer 3                    & \multicolumn{3}{c}{size\,10, softmax}\\  \midrule
\multicolumn{5}{c}{\scshape\gls{convnet} parameters}\\ \midrule
$\mat W_\mathrm{init}$, $\vv b_\mathrm{init}$ & initialization     & \multicolumn{3}{c}{uniform Kaiming He\cite{he2016deep}}      \\ 
                                              & batchsize                     & \multicolumn{3}{c}{64}\\
$\mathcal L$                                  & loss function                 & \multicolumn{3}{c}{cross-entropy}\\
                                              & optimizer                     & \multicolumn{3}{c}{stochastic gradient descent}\\
$\eta_{\mat W, \vv b}$                         & learning rate                 & \multicolumn{3}{c}{0.01}\\
$\mu$                                         & momentum                      & \multicolumn{3}{c}{0.9} \\ \midrule
\multicolumn{5}{c}{\scshape symmetrization parameters} \\ \midrule
$\mat B_\mathrm{init}$  & initialization     & \multicolumn{3}{c}{uniform Kaiming He\tnote{a}}     \\ \midrule
\multicolumn{5}{c}{{\scshape KP}\tnote{b}} \\ \midrule
$\lambda$  & weight decay factor & \multicolumn{3}{c}{0.001} \\ \midrule
\multicolumn{5}{c}{{\scshape \gls{salnet}}\tnote{b}} \\ \midrule
   & layer sizes & \multicolumn{2}{c}{\numlist{1600;384;192;10}} & \numlist{1024;384;192;10} \\
   & neuron model & \multicolumn{3}{c}{\gls{glm} with rectangular \gls{psp}} \\
   simulation time step  &  \multicolumn{2}{c}{\SI{1}{\milli\second}}  \\
\tauref  & refractory period & \multicolumn{3}{c}{\SI{10}{\milli\second}} \\
\tausyn  & synaptic time constant & \multicolumn{3}{c}{\SI{10}{\milli\second}} \\
$\eta_\mathrm{SAL}$ & \gls{sal} learning rate & \multicolumn{3}{c}{0.04} \\
  & number of updates per epoch & \multicolumn{3}{c}{5}  \\
  & duration per update & \multicolumn{3}{c}{\SI{64}{\second}} \\ \midrule
\multicolumn{5}{c}{{\scshape \gls{rddnet}}\tnote{b}\tnote{c}} \\ \midrule
   & layer sizes & \multicolumn{2}{c}{\numlist{1600;384;192;10}} & \numlist{1024;384;192;10} \\
   & neuron model & \multicolumn{3}{c}{\gls{lif} with \gls{rdd} stimulation protocol} \\
$T$ & total stimulation time & \multicolumn{3}{c}{\SI{90}{\second}} \\
\\\bottomrule                                                                                                                       
\end{tabularx}

\begin{tablenotes}[flushleft]  %
  \small
  \item[a] Not applicable to \gls{bp}, because there are no explicit backward weights.
  \item[b]\label{tn:kp} Only applied to \gls{fc} layers 1\,--\,3. The convolutional layers are trained with \gls{fa}.
  \item[c] All \gls{rdd}-specific model parameters are the same as in the original publication~\cite{guerguiev2019spikebased}.
\end{tablenotes}

\end{threeparttable}

\end{table*}

\subsection{SAL and Dale's law}
\label{sec:methods:dales_law}

The setup for the experiment in \cref{sec:results:dales_law} consists of two excitatory (indicated by a superscript $\cdot^E$) and two inhibitory neurons ($\cdot^I$), each modeled by a \gls{glm} (\cref{eq:methods:mempot_gls}).
Each excitatory neuron projects through a direct pathway to the other excitatory neuron via $W^{EE}_{ij} > 0$ and through an indirect pathway first to an inhibitory neuron via $W^{IE}_{ii} > 0$ and this inhibitory neuron connects via $W^{EI}_{ij} < 0$ to the other excitatory neuron.

The biases $b^I_i$ are set to high negative and the weights $W^{IE}_{ij}$ to high positive values, to let the inhibitory neurons behave like statistical parrot neurons.
To account for neuronal variability in real physical networks, we add a random number $\xi \sim \mathcal{N}(0, \sigma=0.2)$ drawn from a Gaussian distribution to all weights and biases.

We are interested in the effective weight $W^\mathrm{eff}_{ij}$, i.e., the total effect of input from the direct and indirect pathway.
This can be estimated by 
\begin{equation}\label{eq:methods:w_eff}
    W^\mathrm{eff}_{ij} = W^{EE}_{ij} + W^{EI}_{ij} \frac{\sigma \left(W^{IE}_{jj} r^E_j + b^I_j\right)}{r^E_j} \;.
\end{equation}
While this formula is exact for purely rate-based models, it can be used as an approximation for spiking systems.
In \cref{fig:results:ei_eff_weights}\crefanno{b}, we compare the output rates of excitatory neuron 2 receiving input through strictly excitatory and inhibitory pathways with an equivalent connection $W^\mathrm{eff}_{ij}$ (violating Dale's law).
We observe a close correspondence of \cref{eq:methods:w_eff} with the actual effective weight.

the other weights, in particular in the indirect pathway, are constant.
However, learning $W^{EI}_{ij}$ while keeping $W^{EE}_{ij}$ fixed yields qualitatively the same results.
In the \gls{ppd} in \cref{fig:results:ei_eff_weights}\crefanno{c}, all data points are simulated numerically to compute the \gls{sal} updates (as opposed to \cref{fig:results:sal}\crefanno{e} and \cref{fig:results:psps}, which are computed with the method in \cref{sec:methods:analytic_stdd}).

\begin{table}[htb]
\caption{
    Simulation parameters for the Dale's law compatible two neuron system.
}
\label{tab:methods:ei_system_params}
\begin{tabularx}{\linewidth}{Xc}
\toprule
                    & Network obeying Dale's law                 \\ \midrule
$W^{EE}_{12,\, 21}$ & can vary between 0 and 2           \\
$W^{IE}_{11,\, 22}$ & $\sim \mathcal N(4, 0.2)$          \\
$W^{IE}_{12,\, 21}$ & $\sim \mathcal N(-1, 0.2)$         \\
$b^E_{1,\, 2}$      & $\sim \mathcal N(0, 0.2)$          \\
$b^I_{1,\, 2}$      & $\sim \mathcal N(-2, 0.2)$         \\
PSP kernel          & rectangular with $\tausyn = \tauref$ \\ \bottomrule
\end{tabularx}
\end{table}

\subsection{Other PSP shapes }
\label{sec:methods:psps}

For calculating the \emph{average relative deviation} between the true attractor and the diagonal $W_{21} = W_{12}$ in the \gls{ppd} \cref{fig:results:psps}, we define a map $W'_{21} \to W'_{12} = g(W'_{21})$, that characterizes all points $\mat W' = (W'_{21}, W'_{12})$ on the attractor.
The function $g$ does not need to be defined or fitted by a analytical function, instead, we determine enough points on the attractors numerically.
It also does not matter if we map $W'_{21} \to W'_{12}$ or vice versa, since the problem is symmetric under the exchange of indices.
We define the relative deviation of a point on the attractor from the diagonal by
\begin{equation}\label{eq:methods:deviation}
    \delta(W_{21}) \coloneqq \frac{W_{21} - g(W_{21})}{W_{21} + g(W_{21})}.
\end{equation}
The average relative deviation is then
\begin{equation}\label{eq:methods:av_deviation}
    D \coloneqq \frac{1}{W_{21}^\mathrm{max} - W_{21}^\mathrm{min}}\int\limits_{W_{21}^\mathrm{min}}^{W_{21}^\mathrm{max}} |\delta(W'_{21})|\, \dd W'_{21},
\end{equation}
where $W_{21}^\mathrm{max}$ and $W_{21}^\mathrm{min}$ are the maximal and minimal weight values in the \gls{ppd}.

\section*{Code availability}

The simulations were performed by custom code written in Python (v3.11), numpy (v2.0) and numba (v0.60). The \gls{symmnet} simulations were done using PyTorch (v2.7).
All code is made available under \url{https://github.com/unibe-cns/sal-code}.

\section*{Data availability}

All data analyzed in this study can be reproduced by executing the code made available under \url{https://github.com/unibe-cns/sal-code}.

\section*{Acknowledgements}

We would like to thank Everton Agnes for an insightful discussion about the plethora of plasticity mechanisms observed in cortex.
This research has received funding from the European Union's Horizon 2020 research and innovation programme under grant agreement Nos. 720270, 785907, and 945539 (Human Brain Project, HBP) and from the European Union’s Horizon Europe Programme under the Specific Grant Agreement No. 101147319 (EBRAINS 2.0 Project).
Some simulations were performed on the bwForCluster NEMO, supported by the state of Baden–Württemberg through bwHPC and the German Research Foundation (DFG) through grant no.~INST 39/963-1 FUGG.
We further acknowledge the use of EBRAINS and Fenix Infrastructure resources, which are partially funded from the European Union's Horizon 2020 research and innovation programme through the ICEI project under the grant agreement No. 800858.
The contribution of KM has received funding by the Volkswagen Foundation and the Swiss National Science Foundation.
Finally, we would like to express our deepest gratitude to the Manfred Stärk Foundation, whose unwavering support has made this, and many other projects possible.

\printbibliography
\end{refsection}

\clearpage

\begin{refsection}
\makeatletter
\gdef\thepage{S\arabic{page}}
\makeatother
\setcounter{page}{1}

\section{Supplementary Material}
\label{sec:sm}

\subsection{Validation of the analytical \gls{stdd} computation}
\label{sec:sm:stdd_ana_vs_sim}

\begin{figure}[htb]
    \centering
    \includegraphics[width=1\linewidth]{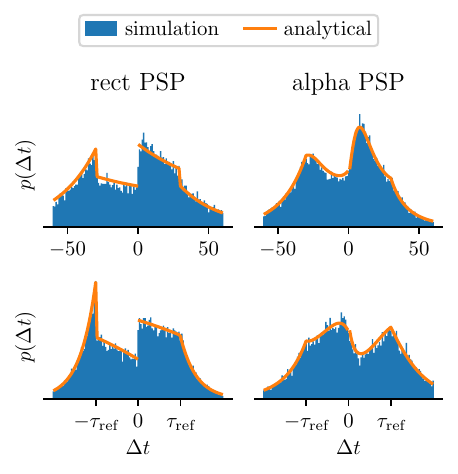}
    \caption{
        \captiontitle{Comparison between the analytical method to compute \glspl{stdd} and numerical simulations.}
    }
    \label{fig:sm:stdd_ana_vs_sim}
\end{figure}

\Cref{fig:sm:stdd_ana_vs_sim} demonstrates that the \glspl{stdd} computed with the method in \cref{sec:methods:analytic_stdd} agrees with a numerical simulation of an equivalent two neuron system.
For each panel, weights and biases are randomly drawn from a uniform distribution $\mathcal U(-1, 1)$.
The time constant of the alpha-shaped PSP is chosen to be $\tausyn = \frac{1}{3} \tauref$, the total simulation time is \num{33000}~$\tauref$.
All four panels show good agreement between the analytical method and the simulation.

\subsection{Analysis of the Kolen-Pollack weight decay rate in spiking sampling networks}
\label{sec:sm:kp_lr}

\begin{figure*}[htb]
    \centering
    \includegraphics[width=1\linewidth]{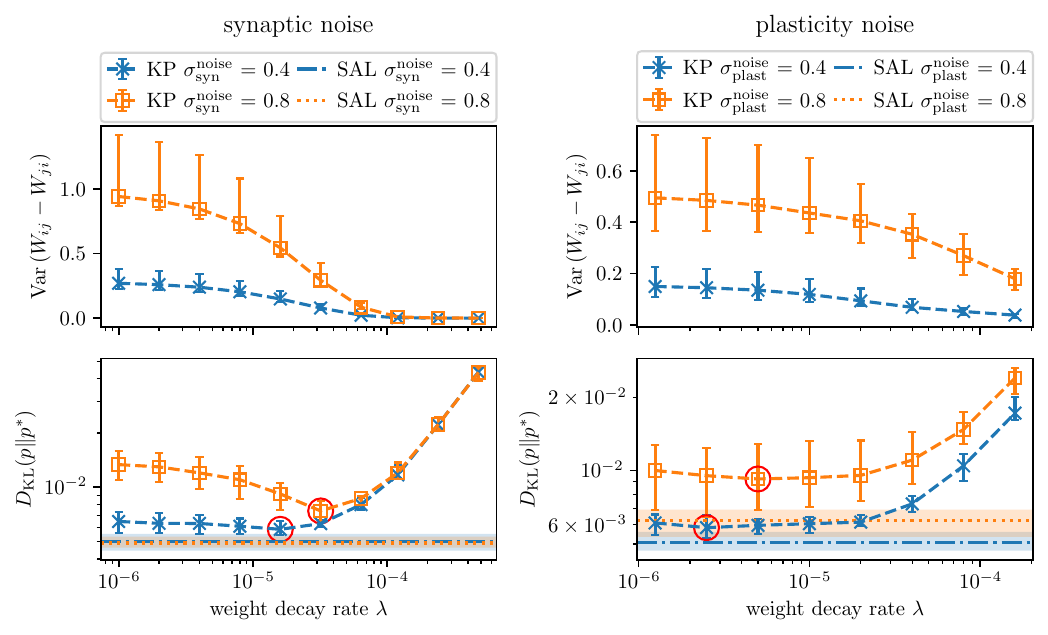}
    \caption{
    \captiontitle{Choosing the optimal Kolen-Pollack weight decay rate $\lambda$ for the \gls{ssn} experiment.}
    Depicted is the final asymmetry and \gls{dkl} after \num{2000} wake-sleep cycles as a function of the weight decay rate $\lambda$.
    The straight lines are the bottom of the lower panels show the \gls{sal} results for reference.
    The lowest \gls{dkl} values are encircled in red.
    }
    \label{fig:sm:kp_lr}
\end{figure*}

In the following, we access the influence of the weight decay rate $\lambda$ on both the symmetrization and the \gls{dkl} in \glspl{ssn} trained with the \glsxtrfull{kp} (\cref{eq:results:kp}).
While a higher $\lambda$ results in better symmetrization, it also interferes with the wake-sleep algorithm itself: 
As performance improves and the weight updates from wake-sleep reach the same magnitude as the weight decay, learning will stall.
Hence, the goal of achieving low \glspl{dkl} demands small decay rates, giving rise to an inevitable performance trade-off between gradient descent and weight symmetrization updates as demonstrated in \cref{fig:sm:kp_lr} (left column).

Moreover, given a finite training time, each noise level has a different optimal $\lambda$ for achieving the lowest \gls{dkl} (compare the minimum for the orange and blue curves in the \gls{dkl} of the synaptic noise scenario).
This would effectively require adapting the weight decay rate to the -- usually unknown -- noise levels of the system.
For the synaptic noise scenario in \cref{fig:results:snn_init_noise}, we choose the $\lambda$ corresponding to the minimum of the $\sigma_\mathrm{syn}^\mathrm{noise} = 0.8$ curve.

The presence of plasticity noise poses another issue:
The standard \gls{kp} algorithm uses identical decay rates on both synapses, which are fundamentally incapable of counterbalancing the asymmetric weight updates originating from noisy plasticity.
Hence, weight decay has little effect on the symmetry, unless the decay rate is so strong that it will suppress the wake-sleep learning (\cref{fig:sm:kp_lr}, right column).
For the plasticity noise scenario in \cref{fig:results:snn_init_noise}, we choose a $\lambda$ of \num{4e-5}, which doesn't deteriorate the wake-sleep learning too much but still has a measurable effect on the weight symmetry.

\subsection{Extended comparison of symmetrization algorithms in deep networks}
\label{sec:sm:symmnet_extended}

For further contextualizing the results of the \gls{symmnet} experiment presented in \cref{fig:results:symmnet}, we additionally train the \gls{convnet} with two more symmetrization algorithms:

\begin{enumerate}
    \item A combination of \gls{bp} in the fully connected layers and feedback alignment in the convolutional layers (BP~+~FA).
    Due to the weight sharing problem in the convolutional layers, any symmetrization algorithm is
    only applied to the fully connected layers only in the \gls{symmnet} architecture.
    Using \gls{bp} in the fully connecting layers models the best achievable symmetrization (namely when $\mat B = \mat W^\transp$ is enforced all the time) and therefore serves as a fair upper limit for any real symmetrization scheme.
    \item One variant of \gls{scfa} given by $\mat B_t = \sgn(\mat W_t) |\mat B_0|$, where $\mat B_0$ is a fixed random backward matrix~\cite{liao2016important}.
    However, this implementation still requires to transport the sign of the forward weight.
\end{enumerate}

The results are shown in \cref{tab:sm:symmnet_extented}.
Among all symmetrization algorithms, \gls{sal} comes closest to the optimal limit given by BP~+~FA due to its ability to best align forward and backward weights.

\begin{table*}[htb]
    \centering
    \caption{Final test accuracies of the \gls{symmnet} experiment, augmented by a combination of \gls{bp} in the fully connected layers and \gls{fa} in the convolutional layers (BP \& FA) as a model for the best possible symmetrization algorithm in this setting and \glsxtrfull{scfa} algorithm.}
    \begin{tabular}{lccccccc}
\toprule
 & BP & BP + FA & FA & SC FA & SAL & KP & RDD \\
\midrule
CIFAR-10 & 83.8$\pm$0.3 & 80.4$\pm$0.3 & 69.5$\pm$0.7 & 78.4$\pm$0.5 & 79.6$\pm$0.4 & 78.7$\pm$1.0 & 78.7$\pm$0.4 \\
FMNIST & 92.1$\pm$0.1 & 91.5$\pm$0.3 & 90.0$\pm$0.1 & 91.4$\pm$0.2 & 91.5$\pm$0.2 & 90.8$\pm$0.3 & 91.5$\pm$0.1 \\
SVHN & 93.5$\pm$0.1 & 91.9$\pm$0.1 & 84.0$\pm$0.5 & 90.5$\pm$0.2 & 91.5$\pm$0.1 & 90.9$\pm$0.1 & 89.2$\pm$0.3 \\
\bottomrule
\end{tabular}

    \label{tab:sm:symmnet_extented}
\end{table*}

\FloatBarrier

\printbibliography[title={Supplementary references}]
\end{refsection}

\end{document}